\documentclass[%
 reprint,
 amsmath,amssymb,
 aps,
]{revtex4-2}
\usepackage{comment}
\usepackage{graphicx}
\usepackage{dcolumn}
\usepackage{bm}
\usepackage{braket} 
\usepackage{enumitem}

\usepackage{xcolor} 
\newcommand{\rev}[1]{\textcolor{black}{#1}}
\begin{document}

\preprint{APS/123-QED}

\title{Opening a gap in \rev{the dispersion of the collective excitations} 
of a driven-dissipative condensate \rev{subject to} an external coherent drive}

\author{ E. Stazzu, G. A. P. Sacchetto, I. Carusotto}
\affiliation{INO-CNR Pitaevskii BEC Center and Dipartimento di Fisica, Università di Trento, 38123 Povo, Italy}
\date{\today}

\begin{abstract}
We build a minimal theoretical model to describe the opening of a gap in the dispersion of the collective excitations of a driven-dissipative condensate when the condensate phase is fixed by an additional coherent phase-locking drive. \rev{We map out the phase diagram as a function of the amplitude and frequency of the coherent drive, identifying distinct regions corresponding to different steady-state regimes. For each region, we analyze the dispersion of the collective excitations and determine whether the spectrum is gapped, with either a purely imaginary gap or a finite real part. When the coherent drive is unable to lock the condensate phase, a gapless Goldstone mode is recovered. Within the same phase diagram, we further identify regions of finite-wavevector dynamical instability, where the condensate tends to develop a supersolid-like spatial modulation.} 
While our theoretical framework is directly related to recent experiments with exciton-polariton condensates, it can be applied to describe the effect of external injection also in a variety of spatially extended optical parametric oscillators or laser devices.
\end{abstract}

\maketitle

\section{\label{sec:level1}Introduction} 

The concept of collective excitations is one of the most powerful tools to understand and characterize the physics of many-body states and of the phase transitions connecting them. Originally investigated for weak excitations in conservative systems of material particles at thermal equilibrium, such as electron gases, liquid Helium or dilute Bose-Einstein condensates~\cite{nozieres1999theory,pitaevskii2016bose}, it has recently started receiving a growing interest also in the context of driven-dissipative systems, in particular quantum fluids of light~\cite{carusotto2013quantum} and condensates of photons or polaritons~\cite{bloch2022non}.

The collective excitations of dilute Bose-Einstein condensates of material bosonic particles are accurately described by the Bogoliubov theory~\cite{bogoliubov1947theory}, which predicts an analytical form
\begin{equation}
\omega_B(k)=\sqrt{\frac{\hbar k^2}{2m}\left( \frac{\hbar k^2}{2m} + 2 g n\right) }
\label{eq:Bogo_standard}
\end{equation}
for the dispersion law in terms of the particle mass $m$ and the mean-field interaction energy $g n$ given by the product of the interaction constant $g$ and the particle density $n$. At low-$k$ the dispersion has a sonic-like character $\omega_B(k)\simeq c_s k$ with a speed of sound $c_s=\sqrt{g n/m}$, which transitions to a single-particle-like dispersion $\omega_B(k) \simeq \hbar k^2/(2m)$ at large $k$. The softness $\omega(k\to 0) =0$ of this Bogoliubov  excitation\rev{, referred to as a soft Goldstone mode,} is a direct consequence of the spontaneous breaking of a continuous $U(1)$ symmetry at the condensation phase transition. 
\rev{It is analogous to the soft magnonic branch stemming from the spontaneous breaking of a continuous rotational symmetry in a ferromagnet~\cite{rezende2020fundamentals}.}

The situation is much richer in the case of driven-dissipative systems where the number of particles is not conserved and the steady state originates from a dynamical interplay of pumping and losses, e.g. quantum fluids of light and condensates of photons or polaritons~\cite{carusotto2013quantum,bloch2022non}. As a result of the driven-dissipative condition, a much wider variety of dispersion relations can be observed depending on the specific pumping configuration adopted. 


\rev{In particular, whenever the continuous $U(1)$ symmetry associated to the condensate phase is spontaneously broken, a non-equilibrium generalization of the Goldstone theorem} guarantees the presence of a gapless branch with $\omega(k\to 0)= 0$ in both real and imaginary parts. This occurs in polariton or photon condensates \rev{under an incoherent pumping}, but also in optical parametric oscillators or in generic laser devices~\cite{bloch2022non}. \rev{For driven-dissipative condensates,} several theoretical works~\cite{Wouters:PRB2006,Szymanska:PRL2006,Wouters:PRA2007} have anticipated a diffusive nature of the gapless Goldstone branch at low-$k$,
\begin{equation}
    \omega(k)\simeq -i \alpha k^2\,.
\end{equation}
with a real and positive diffusion coefficient $\alpha$.
An experimental verification of this prediction has been recently reported in~\cite{claude2025observation} using an exciton-polariton condensate in a parametric pumping configuration. On top of this, \rev{in the same experiment} the opening of a gap in the dispersion of collective excitations was reported when the $U(1)$ symmetry is explicitly broken and the condensate phase is externally fixed. 
In contrast to condensates of material particles~\cite{Gunton:PR1968}, this phase-locking can be realized in the optical context by shining an additional coherent phase-fixing beam at a frequency and wavevector in the vicinity of the condensate ones. 
\rev{In the analogy with ferromagnets, this phase-locking corresponds to the pinning of the magnetization in the direction of an external magnetic field, which results in the opening of a gap in the dispersion relation of magnons~\cite{rezende2020fundamentals}. Contrary to standard condensates and ferromagnets at equilibrium, however, the driven-dissipative nature of our optical system has the remarkable consequence that the gap may open in either the imaginary part only, or simultaneously in both the real and the imaginary parts, depending on the details of the configuration. }

Throughout this article we will adopt the terminology of non-equilibrium condensates, but the reader should keep in mind that the results directly extend to the collective excitation modes of spatially extended laser devices and optical parametric oscillators. Also in these contexts, a large literature has addressed the issue of phase locking of an oscillator to an external coherent field~\cite{siegman1986lasers,Adler1946ASO,Paciorek1965,stover1966locking} but investigations of collective modes have only been reported for the simplest few-mode geometries~\cite{Huard:PRApp2019}.

The goal of this work is to extend the generic theory of non-equilibrium condensates~\cite{Wouters:PRL2007} and develop a simple theoretical model of the dispersion of the collective excitations of driven-dissipative condensates in the presence of an additional coherent phase-fixing drive. This theory is then used to draw a phase diagram of the phase-locking process as a function of the frequency and amplitude of the phase-fixing drive. When the phase-fixing is not effective, the dispersion relations keeps displaying a soft Goldstone mode. Within the region of efficient phase-locking, parameter domains are identified where the gap opens either in the imaginary part only or in both the real and imaginary parts of the dispersion. In spite of the simplicity of the model, these results provide an intuitive explanation of the experimental observations in~\cite{claude2025observation}.

In specific, Sec.\ref{sec:Model} introduces the theoretical model and the generalized Bogoliubov formalism to describe the collective excitations around a stationary state or a limit-cycle solution. The physics of different cases of growing complexity is then discussed in the following sections: starting from the non-interacting, zero-detuning case of Sec.\ref{sec:Case1}, the full phenomenology gets visible as soon as a detuning is introduced in Sec.\ref{sec:Case2}. The \rev{effects} of a two-particle interaction term describing the $\chi^{(3)}$ optical nonlinearity of the cavity medium are sketched in Sec.\ref{sec:Case3}. Conclusions are finally drawn in Sec.\ref{sec:Conclu}. Two Appendices report additional details on the analytical calculations.

\section{The model}
\label{sec:Model}

In this Section we lay down the basic theoretical concepts that will be used for the description of the steady-state of the system and of its collective excitations. After a brief review of the standard theory in Sec.\ref{sec:gGPE} and \ref{sec:stationary}, Sec.\ref{sec:BogoFloquet} extends the concept of collective excitations, usually formulated in the literature for the case of stationary state solutions, to steady-states in the form of a limit-cycle. Our choice for the normalization of the different quantities is summarized in Sec.\ref{sec:units}.

\subsection{The generalized Gross-Pitaevskii equation}
\label{sec:gGPE}

A generic theoretical model of the impact of an additional coherent beam on non-equilibrium condensation in a spatially extended planar geometry can be obtained by combining the theories developed in Refs.~\onlinecite{Wouters:PRL2007} and~\onlinecite{Carusotto:PRL2004} for respectively the incoherent and coherent pumping schemes. This leads to a classical field equation for the in-cavity field $E(\textbf{r},t)$ in the form:
\begin{align}
    i\frac{\partial{E}}{\partial{t}} =\quad  &\omega_0 E- \frac{\hbar}{2m^*}\nabla^2 E + g|E|^2E + \notag\\
    &+ \frac{i}{2}\left(\frac{P}{1+ |E|^2/n\rev{_{sat}}}-\gamma\right)E + iE\rev{_{coh}}e^{-i \omega\rev{_{coh}}t}\,.
    \label{starting field equation}
\end{align}
Here, $\omega_0$ is the resonance frequency of the planar cavity and $m^*$ is the effective photon mass. Spatial derivatives are taken along the $\{x, y\}$ in-cavity directions only, while the field along $z$ is considered to be frozen in the lowest cavity mode. The non-linear term proportional to the interaction constant $g$ describes the shift of the optical mode due to a $\chi^{(3)}$ susceptibility of the cavity material and/or exciton-exciton interactions. 
The terms on the second line describe pumping and dissipation: $\gamma$ is the linear loss rate, $P$ is the strength of the incoherent pump and $n\rev{_{sat}}$ is the gain saturation density. \rev{We describe the coherent drive within a rotating-wave approximation and we take it to} be monochromatic and at normal incidence, with a spatially constant amplitude $E\rev{_{coh}}$ and a frequency $\omega\rev{_{coh}}$. In this work, we will indicate the field equation \eqref{starting field equation} as a generalized Gross-Pitaevskii equation \rev{(GPE)} describing the dynamics of a non-equilibrium condensate. In other contexts, very similar equations go under the name of Lugiato-Lefever equation~\cite{columbo2021unifying,Lugiato:Varenna} or Complex Ginzburg-Landau equation~\cite{Aranson:RMP2002}.

For analytical convenience, it is useful to rewrite the field equation \eqref{starting field equation} in a rotating frame at $\omega\rev{_{coh}}$, so to remove any explicit time-dependence from the evolution equation. This leads to an equation for the slowly varying field $\bar{E}(\textbf{r},t) = E(\textbf{r},t)e^{i \omega\rev{_{coh}} t}$ in the form
\begin{align}
    i\frac{\partial{\bar{E}}}{\partial{t}} = \quad &-\Delta\, \bar{E}- \frac{\hbar}{2m^*}\nabla^2 \bar{E}+ g|\bar{E}|^2 \bar{E} + \notag \\ &+ \frac{i}{2}\left(\frac{P}{1+ |\bar{E}|^2/n\rev{_{sat}}}-\gamma\right)\bar{E} + iE\rev{_{coh}}
    \label{rotating frame field equation}
\end{align}
where we have defined $\Delta=\omega\rev{_{coh}}-\omega_{0}$ as the detuning between the coherent drive and the resonant cavity. With no loss of generality, we assume in the following that $E\rev{_{coh}}$ is real-valued and positive. 

\subsection{Stationary states and \rev{Bogoliubov dispersion} of the collective excitations}
\label{sec:stationary}

As a first step, we search for steady-state solutions where the slowly-varying field is stationary and has the same spatial form as the $k=0$ coherent drive, $\bar{E}(\textbf{r},t)=E_{ss}$. This form corresponds to a physical field oscillating at $\omega\rev{_{coh}}$ and locked in phase to the incident field. In the following, we will call this regime as {\em phase-locked} regime.

The stationary state condition leads to an algebraic equation for $E_{ss}$:
\begin{align}
    &\left[\Delta -g|E_{ss}|^2\right] E_{ss}-\frac{i}{2}\left(\frac{P}{1+|E_{ss}|^2/n\rev{_{sat}}}-\gamma\right)E_{ss}=iE\rev{_{coh}}
    \label{steady state equation}
\end{align}
As we will show explicitly in the following sections, this equation can be rearranged to write the intensity \rev{of the coherent incident field} $|E\rev{_{coh}}|^2$ as a function of the stationary state intensity $|E_{ss}|^2$. This formulation will be specially useful to highlight the presence of multi-solution regimes.

The next step consists of studying the collective excitation modes around the stationary state solutions found by solving (\ref{steady state equation}). To this purpose, we consider the ansatz $\bar{E}(\textbf{r},t)=E_{ss}+\delta E(\textbf{r},t)$, where $\delta E(\textbf{r},t)$ is a small spatio-temporally-varying perturbation around the stationary state $E_{ss}$, and we insert it into (\ref{rotating frame field equation}). 

Expanding around $E_{ss}$ and keeping only linear terms in the perturbation $\delta E$, we obtain the following linearized equation of motion
\begin{multline}
     i\frac{\partial}{\partial t}\delta E= \rev{- }\Delta\, \delta E -\frac{\hbar\nabla^2}{2m^*}\delta E+2g|E_{ss}|^2\delta E+\\ +g E_{ss}^2\delta E^*+\frac{i}{2}\left(\frac{P}{1+|E_{ss}|^2/n\rev{_{sat}}}-\gamma\right)\delta E+ \\ -\frac{iP}{2n\rev{_{sat}}(1+|E_{ss}|^2/n\rev{_{sat}})^2}[E_{ss}^2\delta E^*+|E_{ss}|^2\delta E]
    \label{intermediate bogoliubov steps}
 \end{multline}
 that mixes via the nonlinear term the perturbation $\delta E$ with its complex-conjugate $\delta E^*$.

Taking advantage of the translational invariance of the problem under a coherent pump at $k=0$, we can switch to Fourier space and rewrite the equation of motion for the Fourier components $(\delta E_{\textbf{k}}, \delta E_{-\textbf{k}}^*)^T$ in the matrix form:
 \begin{equation}
    i\frac{\partial}{\partial t}
    \begin{pmatrix}
        \delta E_{\textbf{k}} \\
        \delta E_{-\textbf{k}}^*
    \end{pmatrix} = \rev{M_{\textbf{k}}} \begin{pmatrix}
         \delta E_{\textbf{k}}\\
         \delta E_{-\textbf{k}}^*
     \end{pmatrix}
     \label{eq:Bogo_k}
\end{equation}
where
\begin{equation}
    \rev{M_{\textbf{k}}} = \begin{pmatrix}
        a+ib & c \\
        -c^* & -a+ib
    \end{pmatrix}
    \label{eq:matrixM}
\end{equation}
with the short-hands
\begin{eqnarray}
    a&&=-\Delta +\frac{\hbar k^2}{2m^*}+2g|E_{ss}|^2 \notag\\
    b&=&\frac{1}{2}\left(\frac{P}{1+|E_{ss}|^2/n\rev{_{sat}}}-\gamma-\frac{P|E_{ss}|^2}{n\rev{_{sat}}(1+|E_{ss}|^2/n\rev{_{sat}})^2}\right) \notag\\
    c&=&\left(g-\frac{iP}{2n\rev{_{sat}}(1+|E_{ss}|^2/n\rev{_{sat}})^2}\right)E_{ss}^2\notag
\end{eqnarray}
The dispersion relation  as a function of $k$ is then given by the eigenvalues $\omega_{\pm}(k)$ of \rev{$M_{\textbf{k}} $}, which satisfy the equation
\begin{equation*}
    \omega^2_\pm(k)-2ib\omega_{\pm}(k)+|c|^2-a^2-b^2=0
\end{equation*}
By inserting the explicit forms of 
$a$, $b$  and $c$, we obtain the Bogoliubov dispersion relation:
\begin{multline}
    \omega_{\pm}(k) =  \\ = \frac{i}{2}\left(\frac{P}{1+|E_{ss}|^2/n\rev{_{sat}}}-\gamma-\frac{P|E_{ss}|^2} {n\rev{_{sat}}(1+|E_{ss}|^2/n\rev{_{sat}})^2}\right) \rev{+} \\
    \pm \left[ \left(\frac{\hbar k^2}{2m^*}-\Delta +2g|E_{ss}|^2\right)^2 + \right. \\
    \left. -\left( g^2 + \frac{P^2}{4n^2\rev{_{sat}}(1+|E_{ss}|^2/n\rev{_{sat}})^4}\right)|E_{ss}|^4\right]^{1/2} .
    \label{eq: dispersion relations}
\end{multline}
While this expression provides an explicit form of the Bogoliubov dispersion, it depends on the stationary intensity $|E_{ss}|^2$ which must be obtained by solving (\ref{steady state equation}) numerically.

In spite of the formal analogy between this equation and the standard Bogoliubov dispersion in \eqref{eq:Bogo_standard}, a lot of new physics is encoded in the different form of the coefficients \rev{as a consequence of the driven-dissipative nature of the system}.

As a sanity check, one can verify that this form of the dispersion indeed recovers well-known cases available in the literature. 
On one hand, in the absence of incoherent pump $P=0$ the dispersion recovers the one of the coherently pumped fluid~\cite{Carusotto:PRL2004},
\begin{multline}
    \omega_{no-P}(k)= - i\frac{\gamma}{2}+ \\ \pm\sqrt{\left(\frac{\hbar k^2}{2m^*}-\Delta +2g|E_{ss}|^2\right)^2-g^2|E_{ss}|^4}
    \label{eq: bosonic dispersion relations no pump}
\end{multline}
with the various gapped, gapless, and precursor of instability regimes experimentally observed in~\cite{Claude:PRL2022}. On the other hand, in the absence of a coherent pump $E\rev{_{coh}}=0$, the dispersion recovers the diffusive Goldstone mode of a non-equilibrium condensate~\cite{Wouters:PRL2007},
\begin{gather}
    \omega_{no-E\rev{_{coh}}}(k)=-i\frac{\Gamma}{2}\pm\sqrt{\omega_B(k)^2-\frac{\Gamma^2}{4}}
    \label{eq: bosonic dispersion relations no ext field}
\end{gather}
with
\begin{equation}
\Gamma=\gamma \frac{P-\gamma}{P}\,:
\end{equation}
as a consequence of the spontaneously broken $U(1)$ symmetry, the Goldstone theorem guarantees that the dispersion is gapless, i.e. $\omega_{no-E\rev{_{coh}}}(k\to 0)=0$ in both its real and imaginary parts. This has a diffusive behavior at low-$k$, namely $\omega_{no-E\rev{_{coh}}}(k) \simeq -i \alpha k^2$ with positive $\alpha$, giving a zero real part and a quadratically growing imaginary part, as experimentally observed in~\cite{claude2025observation}.

\subsection{Limit cycles and Floquet-Bogoliubov dispersion of the collective excitations}
\label{sec:BogoFloquet}
The stationary solutions discussed so far correspond to configurations in which the condensate is locked in frequency and phase to the incident \rev{coherent} field. But other forms of steady-state  solutions are possible in the late-time limit, in particular closed periodical orbits called \textit{limit cycles}~\cite{kuznetsov1998elements}. In this case, the field ${\bar E}(\textbf{r},t)=E_{ss}^{cyc}(t)$ is spatially uniform but keeps oscillating in time with a period $T$ whose value is not fixed from the outset but is dynamically determined by the evolution and depends on the specific choice of parameters. In terms of the physical field $E(\textbf{r},t)$, this corresponds to a spontaneous oscillation at a dynamically chosen frequency, that is a spontaneous laser oscillation totally {\em unlocked} from the coherent drive. As the $T$-periodic limit cycle $E_{ss}^{cyc}(t)$ is not necessarily purely harmonic and may contain several Fourier components equispaced by $\omega_{ss}=2\pi/T$, the physical emission generally displays a comb of equispaced components at  $\omega\rev{_{coh}}+n\omega_{ss}$~\cite{Huard:PRApp2019}.

To study the collective excitation dispersion around such a limit cycle \rev{in the rotating frame}, we need to linearize (\ref{rotating frame field equation}) for small perturbations around the uniform yet temporally periodic limit-cycle solution,
\begin{equation*}
  \bar{E}(\textbf{r},t)=E_{ss}^{cyc}(t)+\delta E (\textbf{r},t)\,.  
\end{equation*}
As a key difference from the standard Bogoliubov theory, now the zero-order solution is no longer temporally constant but displays a temporal periodicity of period $T$. Instead of considering the linearized evolution in the vicinity of a stationary solution, we thus have to consider it around a given periodic trajectory. 

\rev{To this end, we extend the method developed in~\cite{creffield2009instability} for the excitations of an externally modulated system to the case of a spontaneously-occurring limit cycle. We consider} the linearized propagator $U(T)$ describing the evolution of small perturbations around the limit cycle through a time equal to the period $T$. As for the limit cycle solution we have $E_{ss}^{cyc}(t+T)=E_{ss}^{cyc}(t)$, the linearized propagator $U(T)$ provides a stroboscopic version of the linearized evolution.
The frequencies of the collective excitation modes are then obtained by diagonalizing $U(T)$ and taking the natural logarithm 
\begin{equation}
    \omega_\pm=\frac{i}{T}\,\log\lambda_\pm
\label{eq:log}
\end{equation} 
of its eigenvalues $\lambda_\pm$. While the specific form of the propagator $U(T)$ depends on the initial time $t$ chosen for the Floquet period, its eigenvalues are fully independent of it, giving a well-defined excitation dispersion . However, as typical in Floquet systems~\cite{viebahn2020introduction}, the multi-valued nature of the logarithm makes the dispersion to be defined modulo $\omega_{ss}$: this corresponds to the usual Floquet folding of the bands around the Floquet Brillouin zone of size $\omega_{ss}=2\pi/T$ along the frequency direction. 

As in the stationary case, we will take advantage of invariance under spatial translations to decompose the field in its Fourier components. For each $\textbf{k}$-vector, we then consider the propagator $U_\textbf{k}(T)$ as a $2\times 2$ matrix acting on the $(\delta E_{\textbf{k}}, \delta E_{-\textbf{k}}^*)^T$ components,
\begin{equation}
    \begin{pmatrix}
        \delta E_{\textbf{k}} \\
        \delta E_{-\textbf{k}}^*
    \end{pmatrix}_{t+T} =  U_\textbf{k}(T) \begin{pmatrix}
         \delta E_{\textbf{k}}\\
         \delta E_{-\textbf{k}}^*
     \end{pmatrix}_t\,,
     \label{eq:U_Bogo_k}
\end{equation}
whose eigenvalues provide via \eqref{eq:log} the collective excitation dispersion $\omega_\pm(k)$. \rev{In practice, the propagator $U_\textbf{k}(T)$ can be evaluated as the T-exponential of a time-dependent version of the matrix $M_\textbf{k}(t)$ in \eqref{eq:matrixM} that includes the time-dependence of the limit-cycle in $E_{ss}^{cyc}(t)$.}

\subsection{Units and normalization}\label{sec:units}
For convenience, all figures in this paper and the numerical values reported therein follow the normalization shown in table \ref{tab: normalization}, based on the values of $\gamma$ and $n\rev{_{sat}}$. These parameters correspond, respectively, to the intrinsic loss and the gain saturation, which in a physical system are typically fixed.

\begin{table}[h!]
\centering
\begin{tabular}{ll}
\hline
Quantity & Normalization \\ 
\hline
Intrinsic loss & $\gamma$ \\
Gain saturation & $n\rev{_{sat}}$ \\
Cavity field & $\tilde{E} \equiv \bar{E} / \sqrt{n\rev{_{sat}}}$ \\
Driving field & $\tilde{E}\rev{_{coh}} \equiv E\rev{_{coh}} / \sqrt{n\rev{_{sat}} \gamma^2}$ \\
Incoherent pump & $\tilde{P} \equiv P / \gamma$ \\
Detuning & $\tilde{\Delta} \equiv \Delta / \gamma$ \\
Interaction constant & $\tilde{g} \equiv g \cdot n\rev{_{sat}} / \gamma$ \\
Wavevector & $\tilde{k} \equiv k \cdot \sqrt{\hbar / (2 m^* \gamma)}$ \\
Angular frequency & $\tilde{\omega} \equiv \omega / \gamma$ \\
\hline
\end{tabular}
\caption{\rev{Normalization adopted for plotting the various quantities of the model in the following figures.}}
\label{tab: normalization}
\end{table}

\begin{figure*}[htbp!]
\centering
\begin{minipage}{0.45\linewidth}
    \centering
    \includegraphics[width=\linewidth]{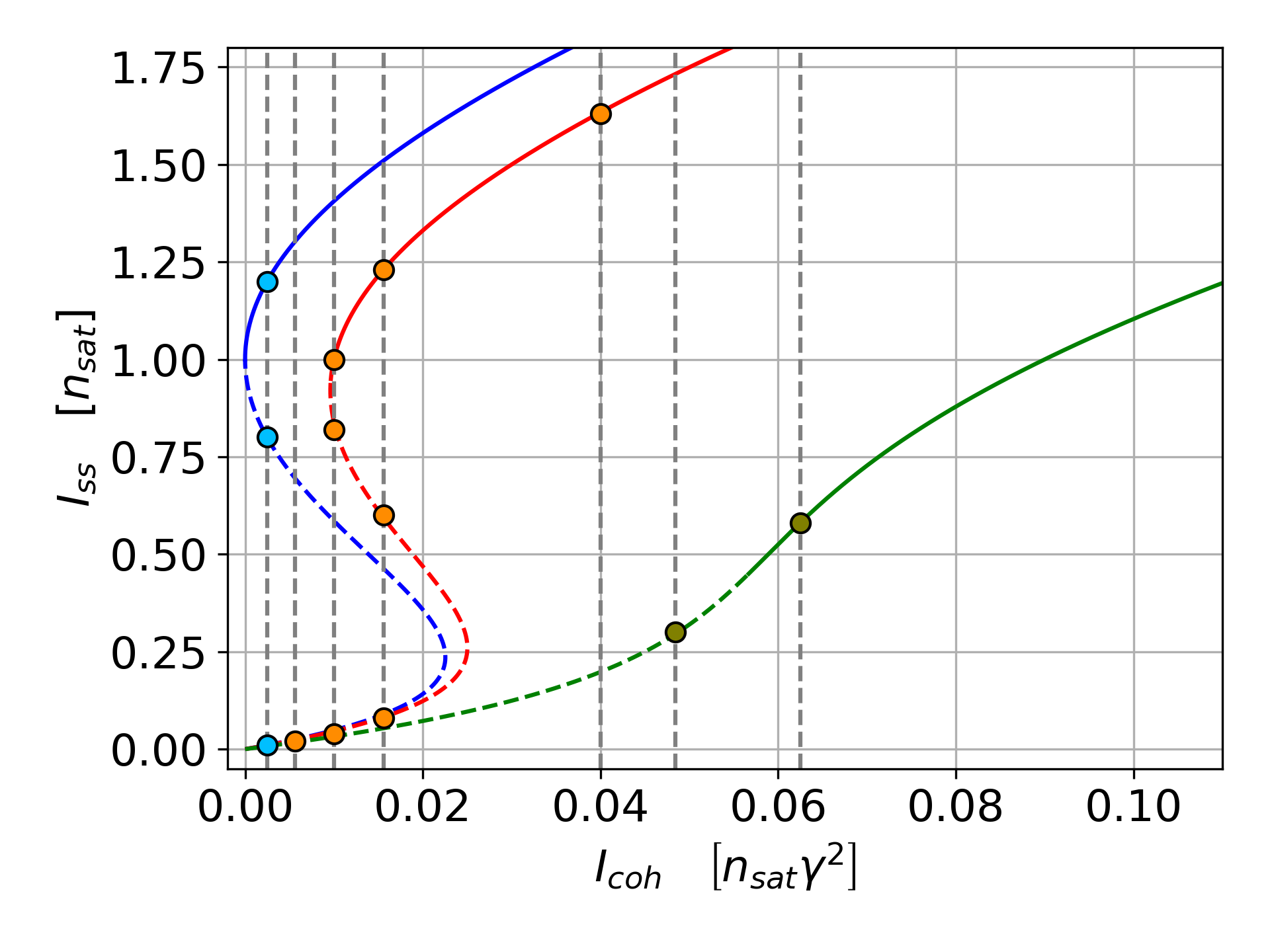}
\end{minipage}\hspace{0.05\linewidth}%
\begin{minipage}{0.43\linewidth}
    \centering
    \includegraphics[width=\linewidth]{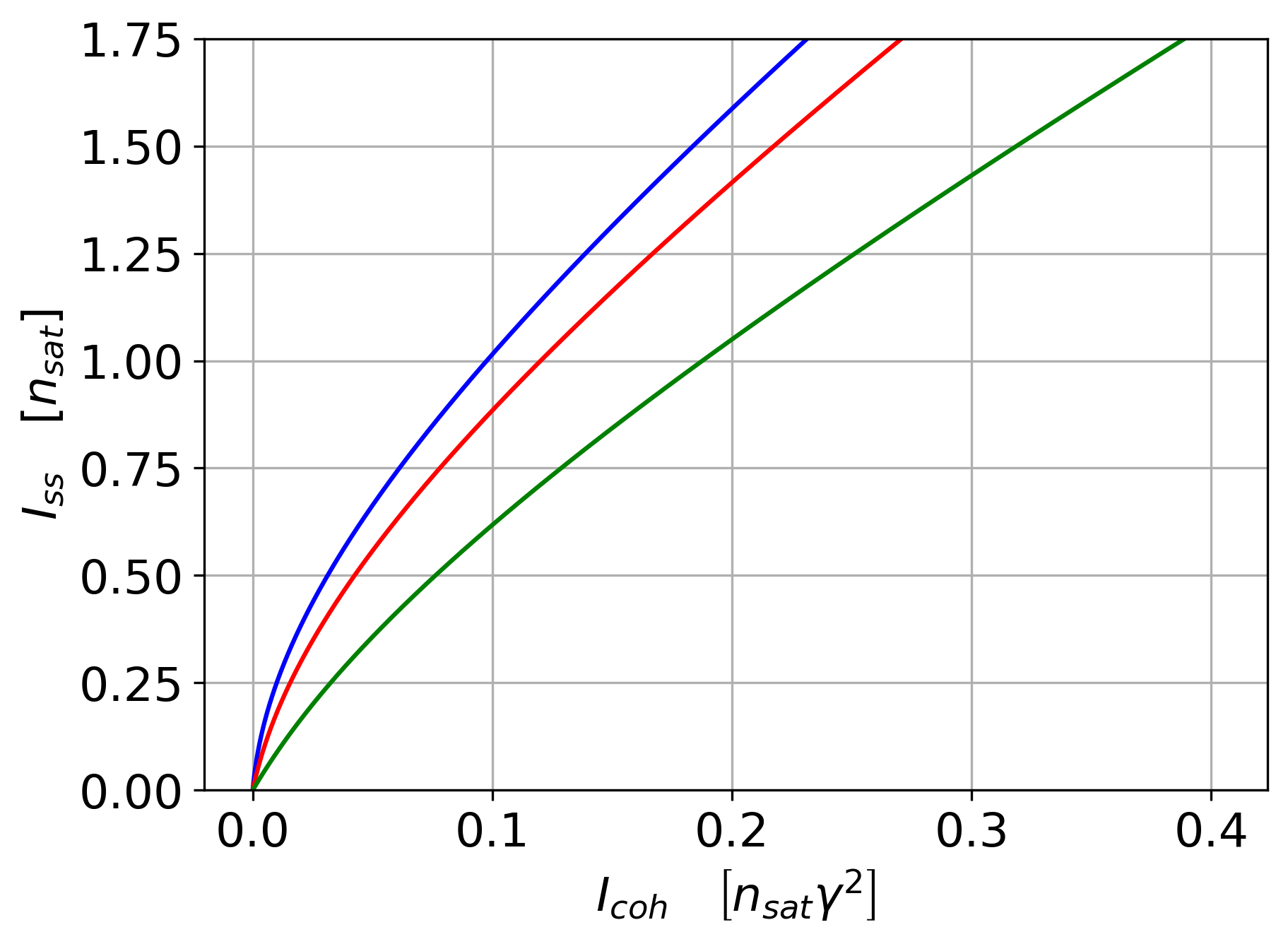}
\end{minipage}
\caption{\rev{Examples of the stationary state intensity $I_{ss}$ vs. incident coherent intensity $I_{\mathrm{coh}}$ for $\tilde{P}=2$ (left) and $\tilde{P}=0.75$ (right), in the absence of interactions ($\tilde{g}=0$). The colored curves correspond to three detuning values: no detuning (blue: $\tilde{\Delta}=0$), weak detuning (red: $\tilde{\Delta}=0.1$), and strong detuning (green: $\tilde{\Delta}=0.3$). For each detuning, solid lines denote dynamically stable spatially uniform steady states at $k=0$, while dashed lines denote dynamically unstable ones. The vertical gray dashed lines indicate the specific values of the coherent intensity at which the numerical vector flow in the complex electric-field plane is computed and shown in Fig.\ref{graph: flow line case 1}-\ref{fig: limit-cycle formation at high Delta}, with the corresponding stationary states highlighted by colored points in the present panel, each selected intensity being applied to a specific detuning case.}
}
\label{graph: qualitative graphs cases 1 and 2}
\end{figure*}

\section{Non-interacting case ($g = 0$)}

In this Section we focus on the non-interacting $g=0$ case for which a comprehensive insight on the different regimes can be obtained with the help of analytical tools. In particular, we will identify the regions of phase locking as a function of the frequency and amplitude of the coherent drive and we will determine the dispersion of the collective excitations in the different regimes. In its relative simplicity, this case already displays most of the basic phenomenology that we will then find also in the general interacting case in the next Section.

\subsection{Resonant drive \rev{$(\Delta=0)$}}
\label{sec:Case1}

As a first, warm-up example, let us focus on the simplest case where the 
coherent field is resonant with the cavity -- \rev{i.e. with detuning $\Delta = \omega_{\rm coh}-\omega_0 = 0$.}

In this case, the equation (\ref{steady state equation}) for the stationary state has the form:
\begin{equation}
    E\rev{_{coh}}=\frac{1}{2}\left(\gamma-\frac{P}{1+|E_{ss}|^2/n\rev{_{sat}}}\right)E_{ss}\,.
    \label{eq: steady state equation case 1}
\end{equation}
Having assumed that $E\rev{_{coh}}$ is real-valued and positive, we conclude from the reality of this equation that $E_{ss}$ must also be real, with a phase difference of either $0$ or $\pi$ with respect to $E\rev{_{coh}}$. This means that the stationary state field is phase-locked to the incident field either in phase or in opposition of phase. Introducing $I_{ss}=|E_{ss}|^2$ and $I\rev{_{coh}}=|E\rev{_{coh}}|^2$ and taking the square modulus of (\ref{eq: steady state equation case 1}), we obtain:
\begin{equation}
    I\rev{_{coh}}=\frac{1}{4}\left(\gamma-\frac{P}{1+I_{ss}/n\rev{_{sat}}}\right)^2I_{ss}
    \label{eq: intesity case 1}
\end{equation}

To identify multi-solution regimes, we study the sign of the derivative $\frac{dI\rev{_{coh}}}{dI_{ss}}$. In fact, when this derivative is negative in some region, the function $I\rev{_{coh}}(I_{ss})$ is no longer monotonically increasing, so  $I_{ss}(I\rev{_{coh}})$ is not a single-valued function but rather shows a multi-valued behavior. Explicit calculation of the derivative gives:
\begin{multline*}
    \frac{dI\rev{_{coh}}}{dI_{ss}} = \frac{1}{4}\left(\gamma-\frac{P}{1+I_{ss}/n\rev{_{sat}}}\right) \times \\ \times\left(\gamma-\frac{P} {1+I_{ss}/n\rev{_{sat}}}+\frac{2PI_{ss}}{n\rev{_{sat}}(1+I_{ss}/n\rev{_{sat}})^2}\right)\,:
\end{multline*}
it is immediate to see that, since $P$, $\gamma$ and $n\rev{_{sat}}$ are all positive, for $P<\gamma$ this expression is always positive, and, therefore, the solution is unique. On the other hand, for $P>\gamma$, the derivative is negative in the interval
\begin{equation}
\frac{n\rev{_{sat}}}{2}\left(\sqrt{\left(\frac{P}{\gamma}\right)^2+8\frac{P}{\gamma}}-2-\frac{P}{\gamma}\right)<I_{ss}<n\rev{_{sat}}\left(\frac{P}{\gamma}-1\right)\,,
    \label{multistable region case 1}
\end{equation}
and the system may exhibit multiple solutions. It is noteworthy that the lower boundary in $I\rev{_{coh}}$ of this multi-solution region is predicted by (\ref{eq: intesity case 1}) to be at $I\rev{_{coh}}=0$: the high-$I_{ss}$ solution then exists 
down to $I\rev{_{coh}}=0$, where it recovers the intensity of the stationary condensate generated by the incoherent pump in the absence of any coherent drive.

Examples of plots of $I_{ss}$ as a function of $I\rev{_{coh}}$ are shown in Fig.\ref{graph: qualitative graphs cases 1 and 2} for the two cases $P<\gamma$ and $P>\gamma$: the existence of multiple stationary solutions for a given $I\rev{_{coh}}$ is visible in this latter case.

\begin{figure*}[htbp!]
\includegraphics[width=0.45\linewidth]{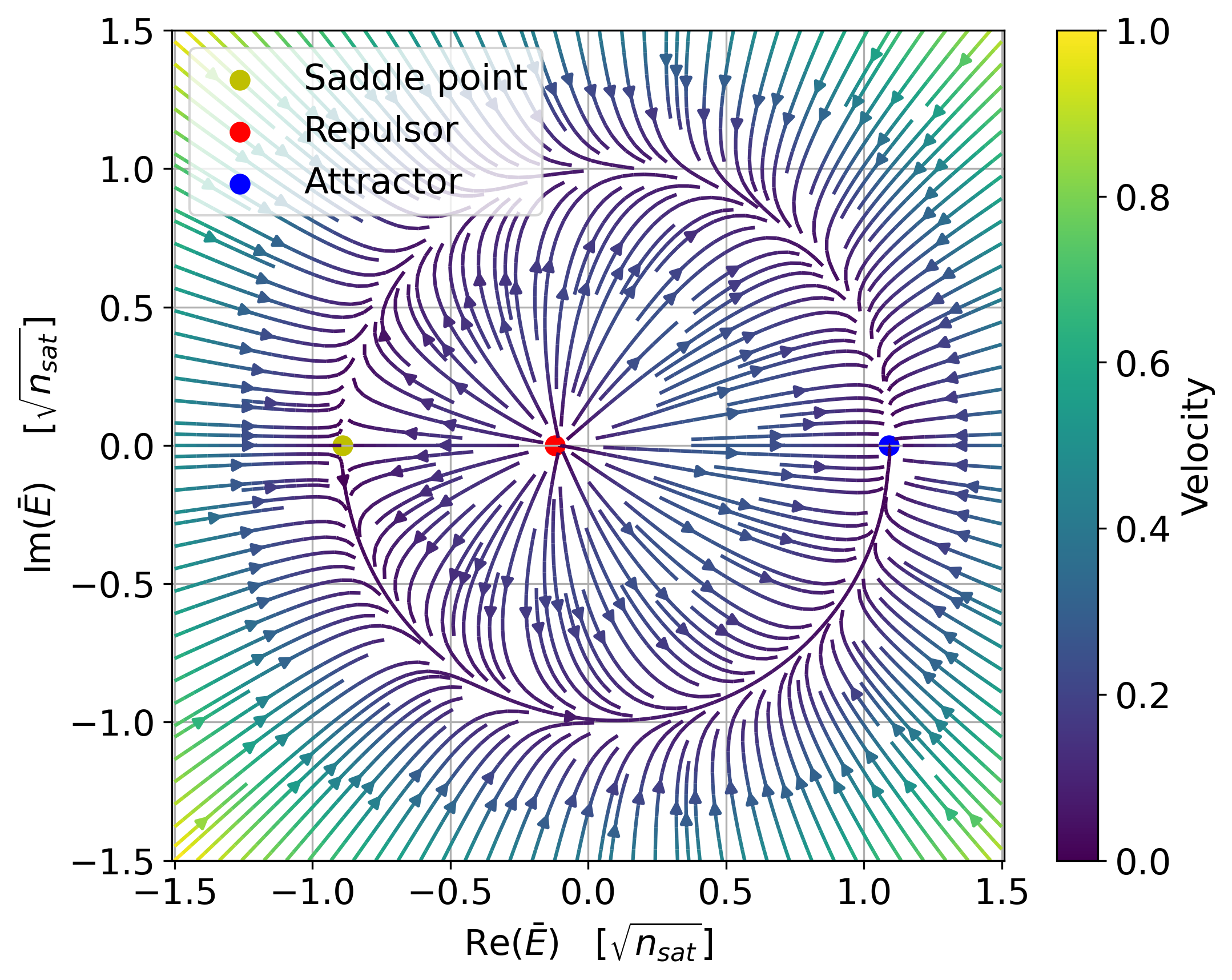}
\includegraphics[width=0.45\linewidth]{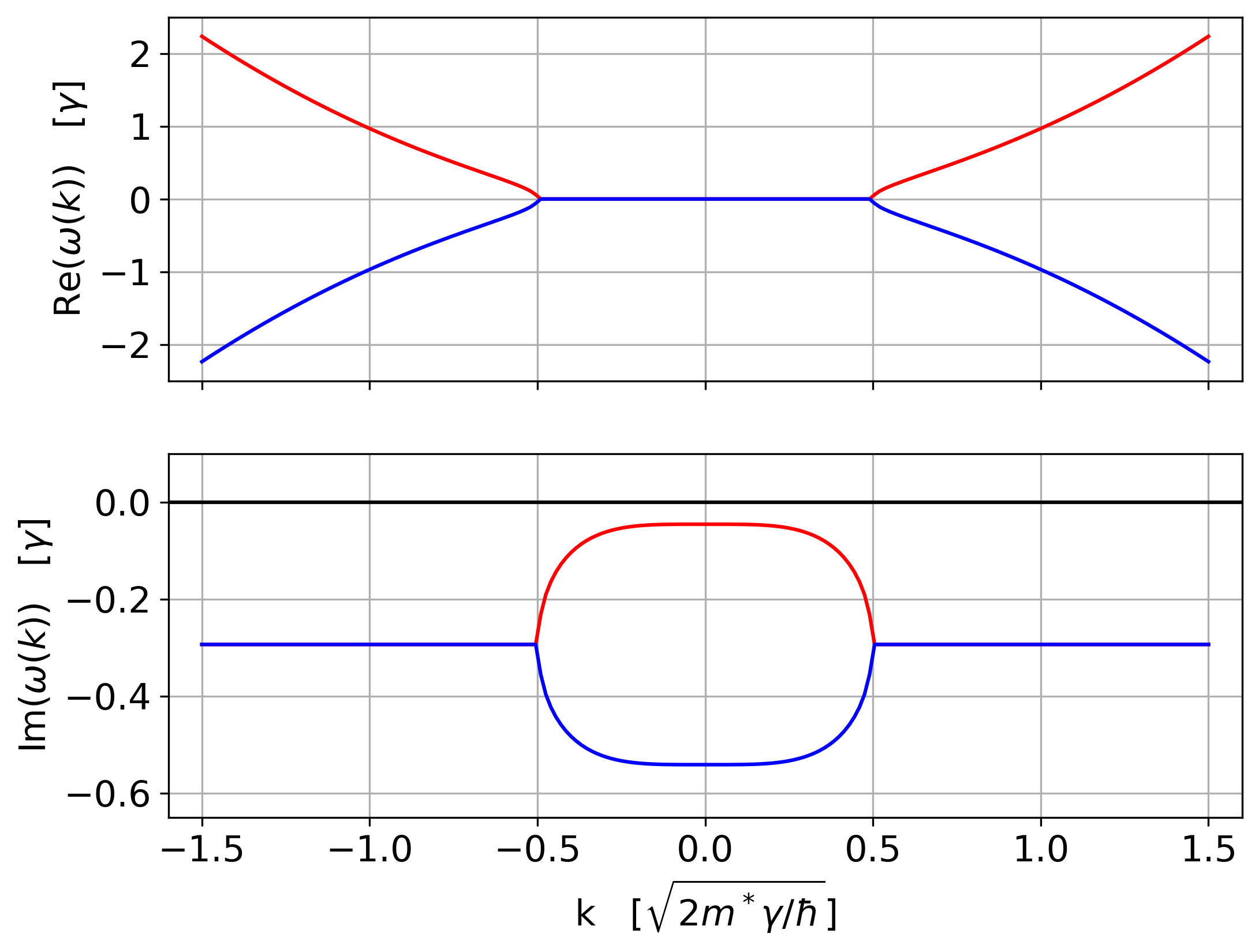}
\caption{\rev{Left: Example of vector flow  of $\bar{E}$ in the complex plane for $\tilde{P}=2$, vanishing detuning ($\tilde{\Delta}=0$) and no interactions ($\tilde{g}=0$), with a weak coherent field $\tilde{E}_{\mathrm{coh}}=0.05$, corresponding to the gray dashed line on the blue curve in the left panel of Fig.\ref{graph: qualitative graphs cases 1 and 2}. Right: Bogoliubov dispersion  for the high-intensity attractor identified in the left panel as the blue point.}}

\label{graph: flow line case 1}
\end{figure*}

To confirm the physical meaningfulness of these solutions, we need to assess their dynamical stability. As a first step in this sense, in the left panel of Fig.\ref{graph: flow line case 1} we show the \rev{vector flow} of \eqref{rotating frame field equation} in the subspace where the field is uniform in space, $\bar{E}(\textbf{r},t)=\bar{E}(t)$. Colored dots indicate the stationary solutions: the two solutions at the lower and intermediate values of $I_{ss}$ with a phase difference $\pi$ from the incident drive turn out to be dynamically unstable, while the highest $I_{ss}$ solution in phase with the drive is dynamically stable. As $I\rev{_{coh}}$ grows out of the multi-solution region, the two unstable lower-$I_{ss}$ solutions merge and disappear, leaving the stable higher-$I_{ss}$ solution unperturbed. In no case for $\Delta=0$ a limit cycle is visible in the flow diagram.\\
\indent While the plot in Fig.\ref{graph: flow line case 1} confirms the stability of the high-$I_{ss}$ solution with respect to spatially-uniform perturbations, a complete study of its stability for generic-$k$ perturbations requires the full Bogoliubov theory of Eq.\eqref{eq:Bogo_k}. As $k$ only enters in the first term in the square-root in \eqref{eq: dispersion relations}, it is straightforward to verify the complete stability of the high-$I_{ss}$ solution. An example of dispersion of the collective excitations around the stable stationary state is shown in the right panel of Fig.\ref{graph: flow line case 1}: the real parts of two branches stick in an interval around $k=0$, while the imaginary parts split. Outside this range, the real part grows in magnitude, eventually approaching the free-particle dispersion. At all $k$ values, however, both branches retain a finite negative imaginary part, which proves overall dynamical stability.

\subsection{General driving frequency \rev{$(\Delta \neq 0$)}}
\label{sec:Case2}

\subsubsection{Steady state: stationary solutions and limit cycles}

For general values of $\Delta$, the equation (\ref{steady state equation}) for the stationary state has the form
\begin{equation}
\centering
    \left[\Delta -\frac{i}{2}\left(\frac{P}{1+|E_{ss}|^2/n\rev{_{sat}}}-\gamma\right)\right]E_{ss}=iE\rev{_{coh}}\,:
    \label{eq: steady state case 2}
\end{equation}
the phase difference between $E_{ss}$ and $E\rev{_{coh}}$ can have arbitrary values
\begin{equation}
    \Delta\phi_{E_{ss}, E\rev{_{coh}}}=\frac{\pi}{2}+\arctan{\left(\frac{1}{2\Delta}\left(\frac{P}{1+|E_{ss}|^2/n\rev{_{sat}}}-\gamma\right)\right)}
\end{equation}
and, by taking the squared modulus of \eqref{eq: steady state case 2}, the relation between the intensities reads
\begin{equation}
    I\rev{_{coh}}=I_{ss}\left(\Delta^2+\frac{1}{4}\left(\frac{P}{1+I_{ss}/n\rev{_{sat}}}-\gamma\right)^2\right)\,.
    \label{eq: intensities case 2}
\end{equation}
Note that changing the sign of $\Delta$ is equivalent to solving the complex conjugate equation: as a result, upon a change in sign of $\Delta$ the field gets conjugated $E_{ss}^*[+\Delta]=E_{ss}[-\Delta]$ but the intensity $I_{ss}$ is identical. Some analytical considerations on the existence of multiple solutions at a given $I\rev{_{coh}}$ in a general $\Delta\neq 0$ case are given in Appendix \ref{app:ss_g=0}. Examples of $I_{ss}$ as a function of $I\rev{_{coh}}$ are shown in Fig.\ref{graph: qualitative graphs cases 1 and 2} for different values of $P/\gamma$ and $\Delta$.

\begin{figure*}[htbp!]
    \centering
    {\includegraphics[width=0.45\linewidth]{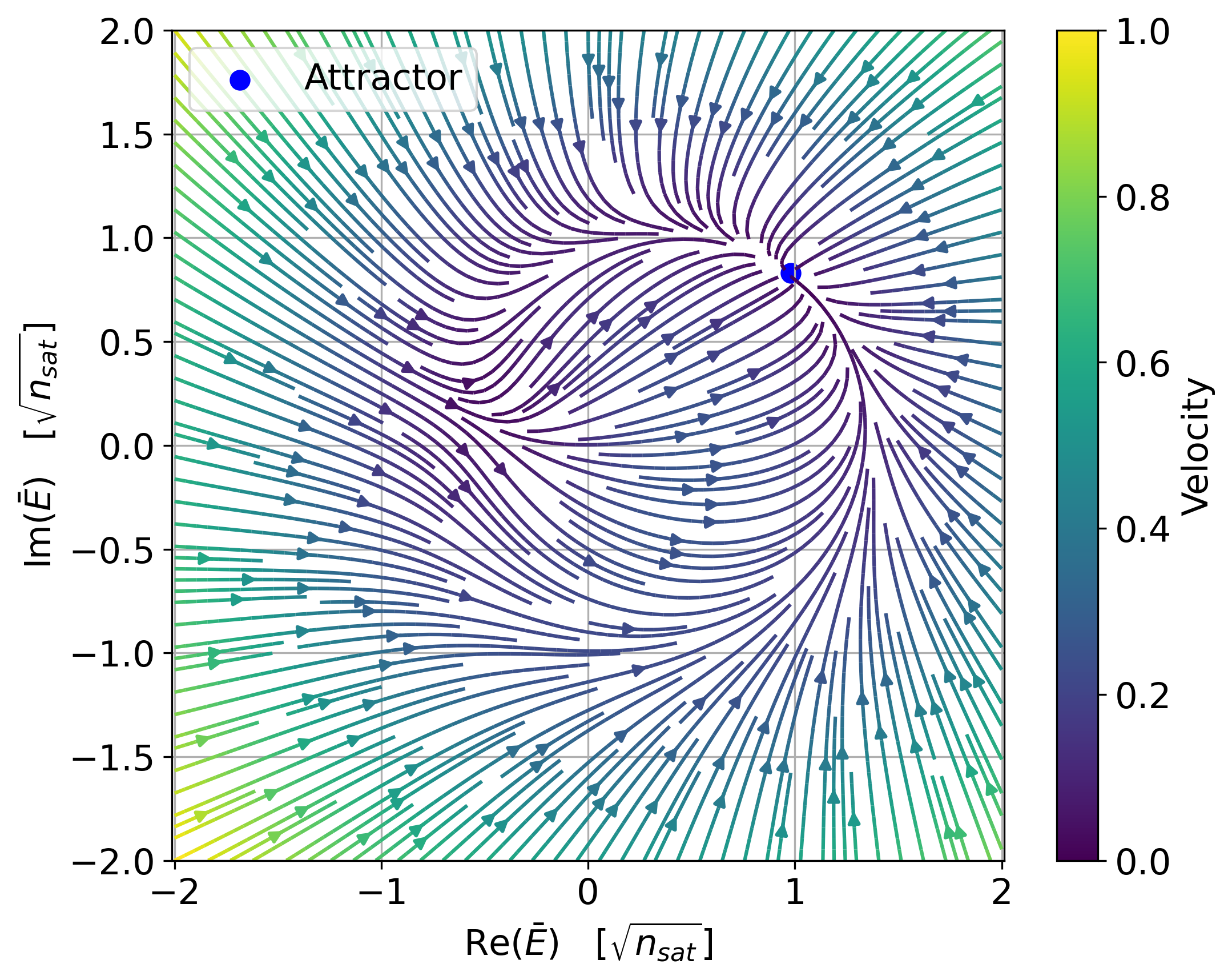}}
    {\includegraphics[width=0.45\linewidth]{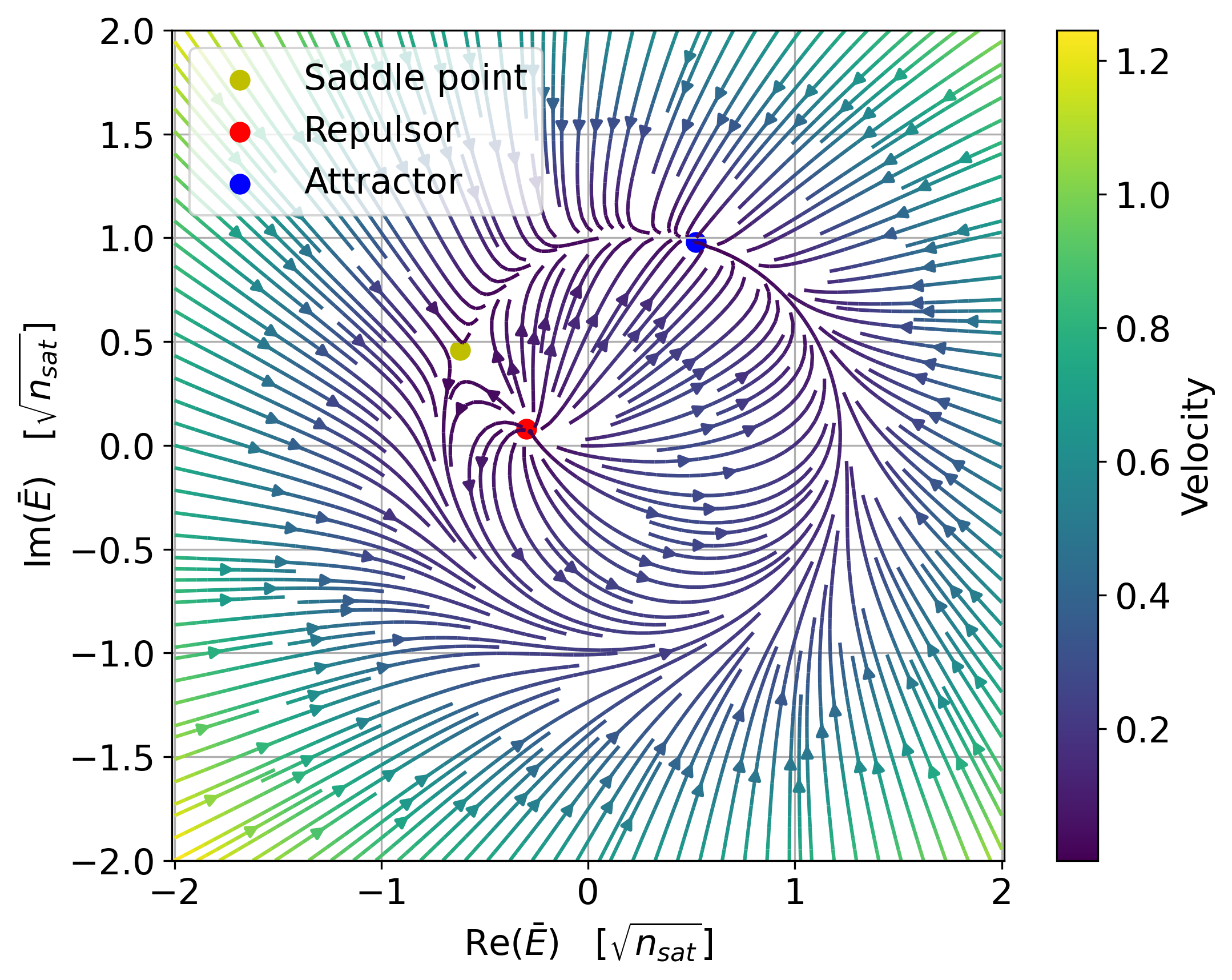}}
    \\
    {\includegraphics[width=0.45\linewidth]{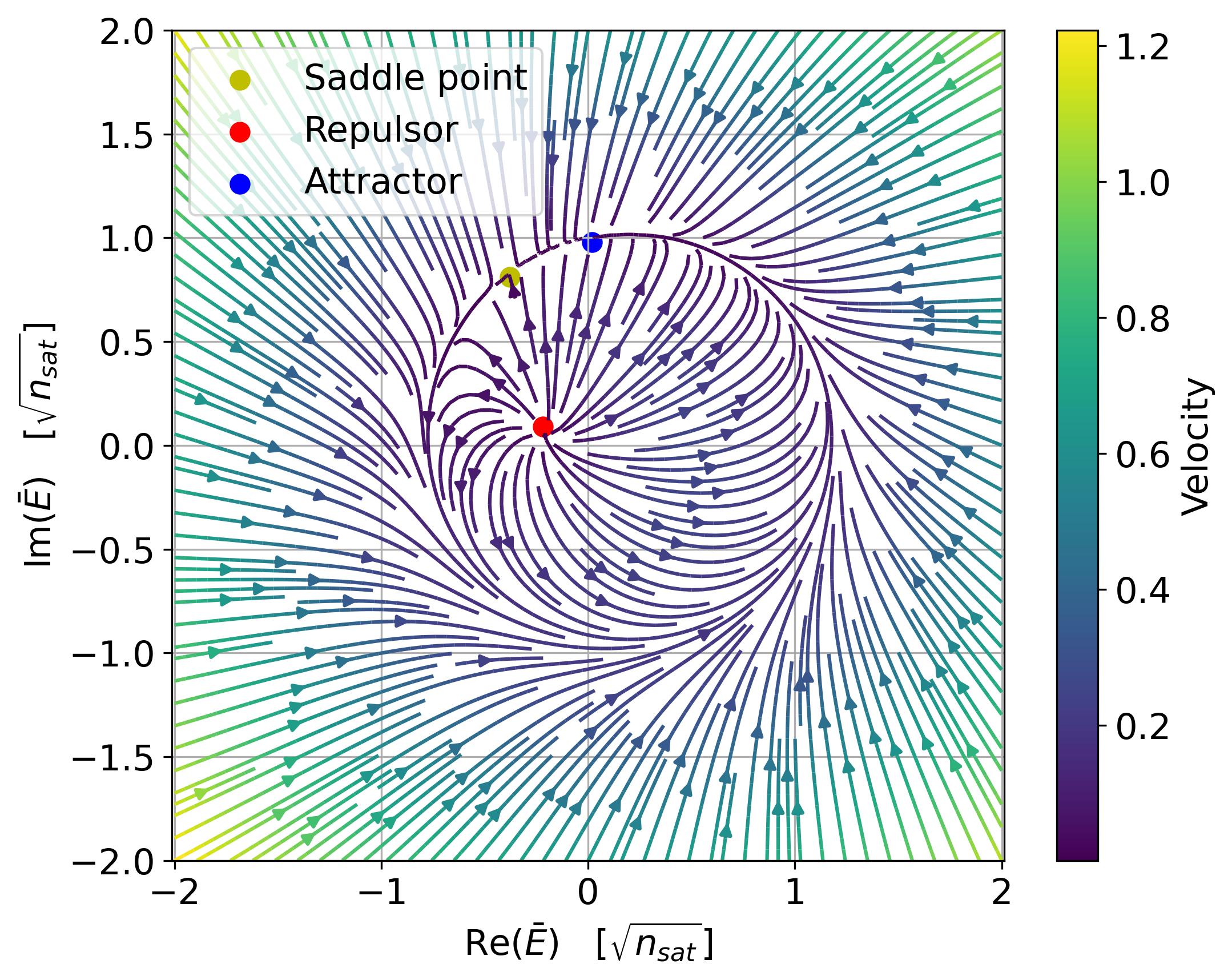}}
    {\includegraphics[width=0.45\linewidth]{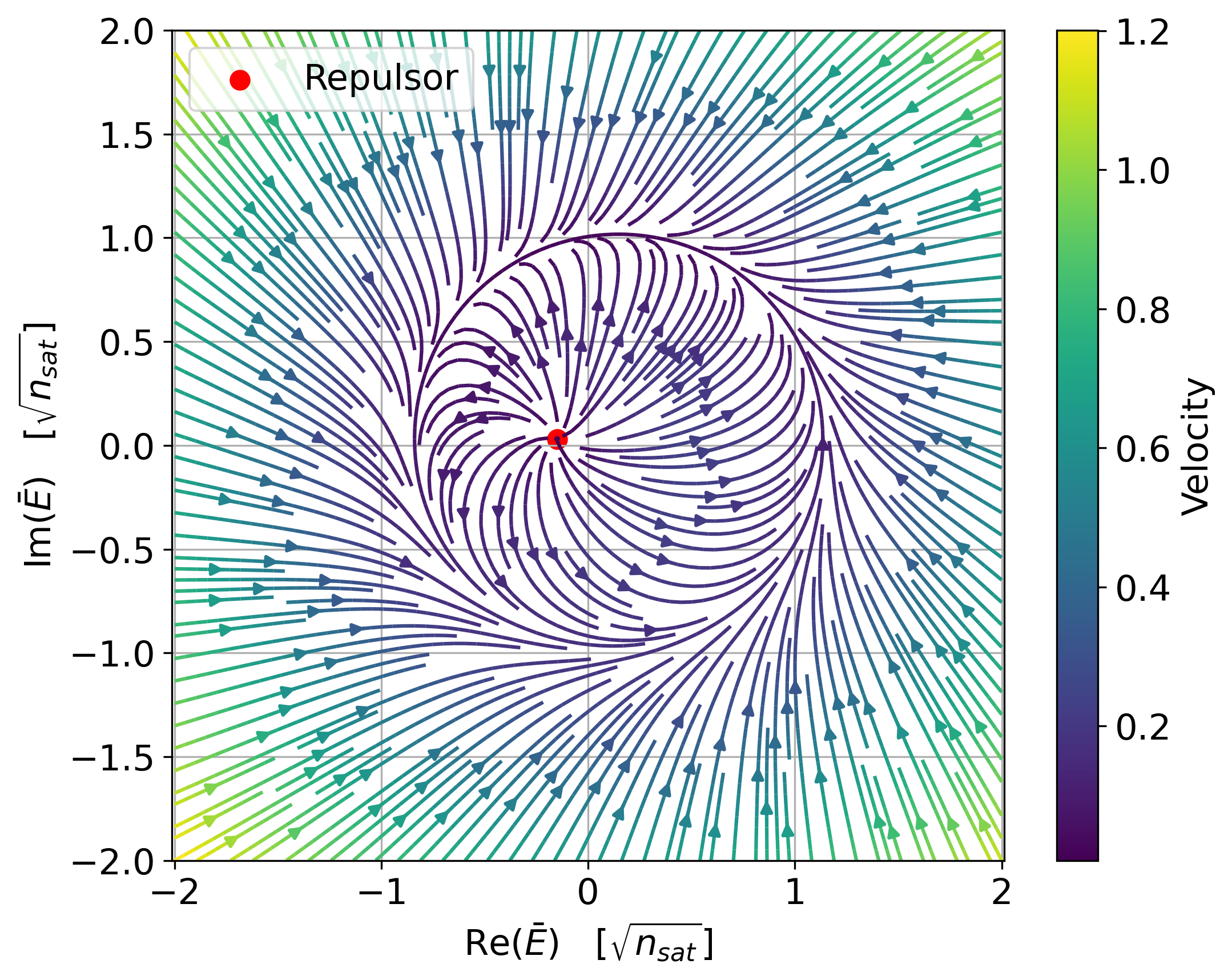}}
    \caption{\rev{Examples of vector flows of $\bar{E}$ in the complex plane \rev{for $\tilde{P}=2$, weak detuning ($\tilde{\Delta}= 0.1$) and no interactions ($\tilde{g}=0$)}, with decreasing values of $\tilde{E}_{coh}= [0.2$, $0.125$, $0.1$, 
$0.075$] corresponding to gray dashed lines on the red curve in the left panel of Fig.\ref{graph: qualitative graphs cases 1 and 2}. The panels show the formation of a limit-cycle from a multi-solution region as $\tilde{E}\rev{_{coh}}$ is decreased. }}
    \label{fig: limit-cycle formation}
\end{figure*}

\begin{figure*}
     \centering
    {\includegraphics[width=0.45\linewidth]{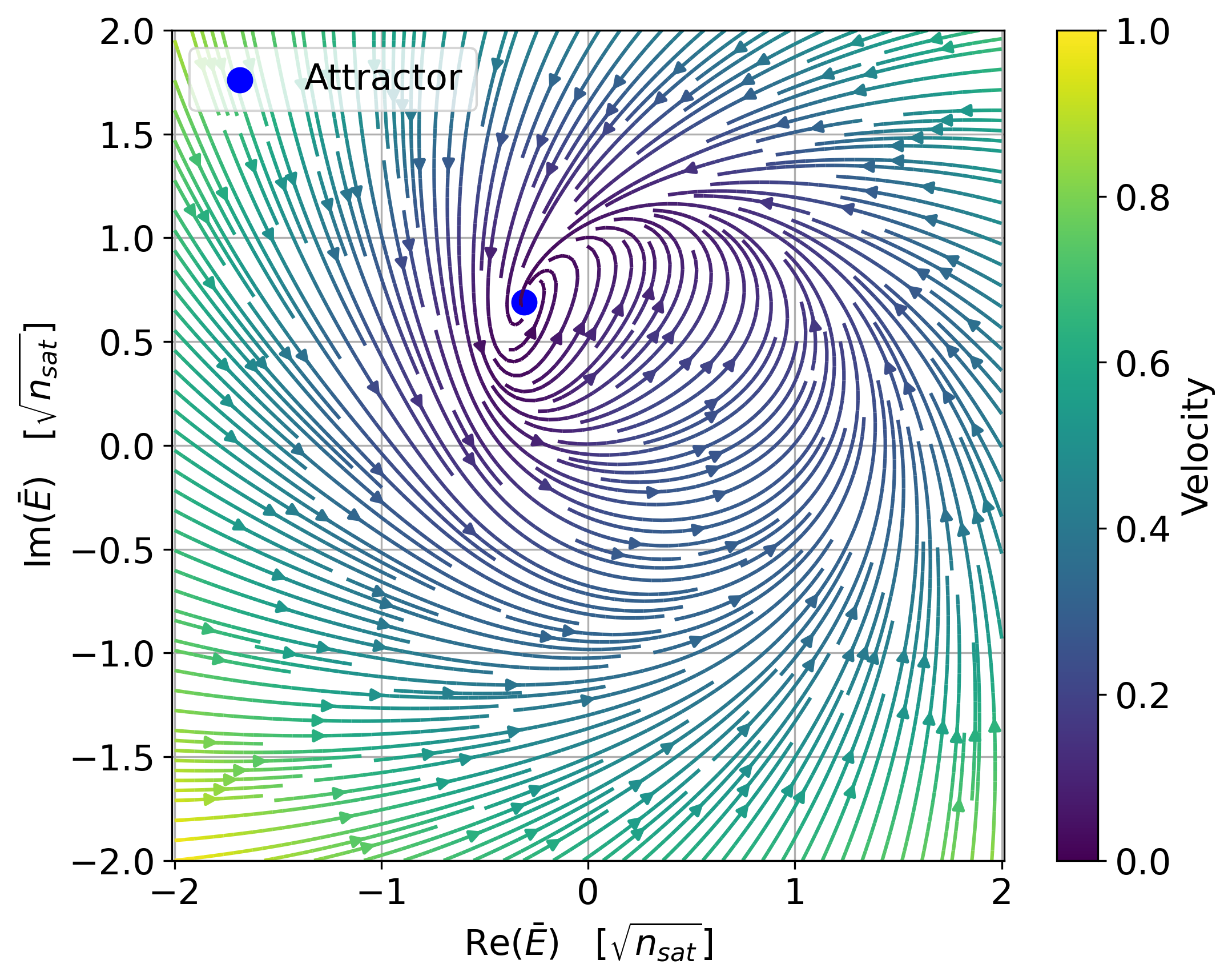}}
    {\includegraphics[width=0.45\linewidth]{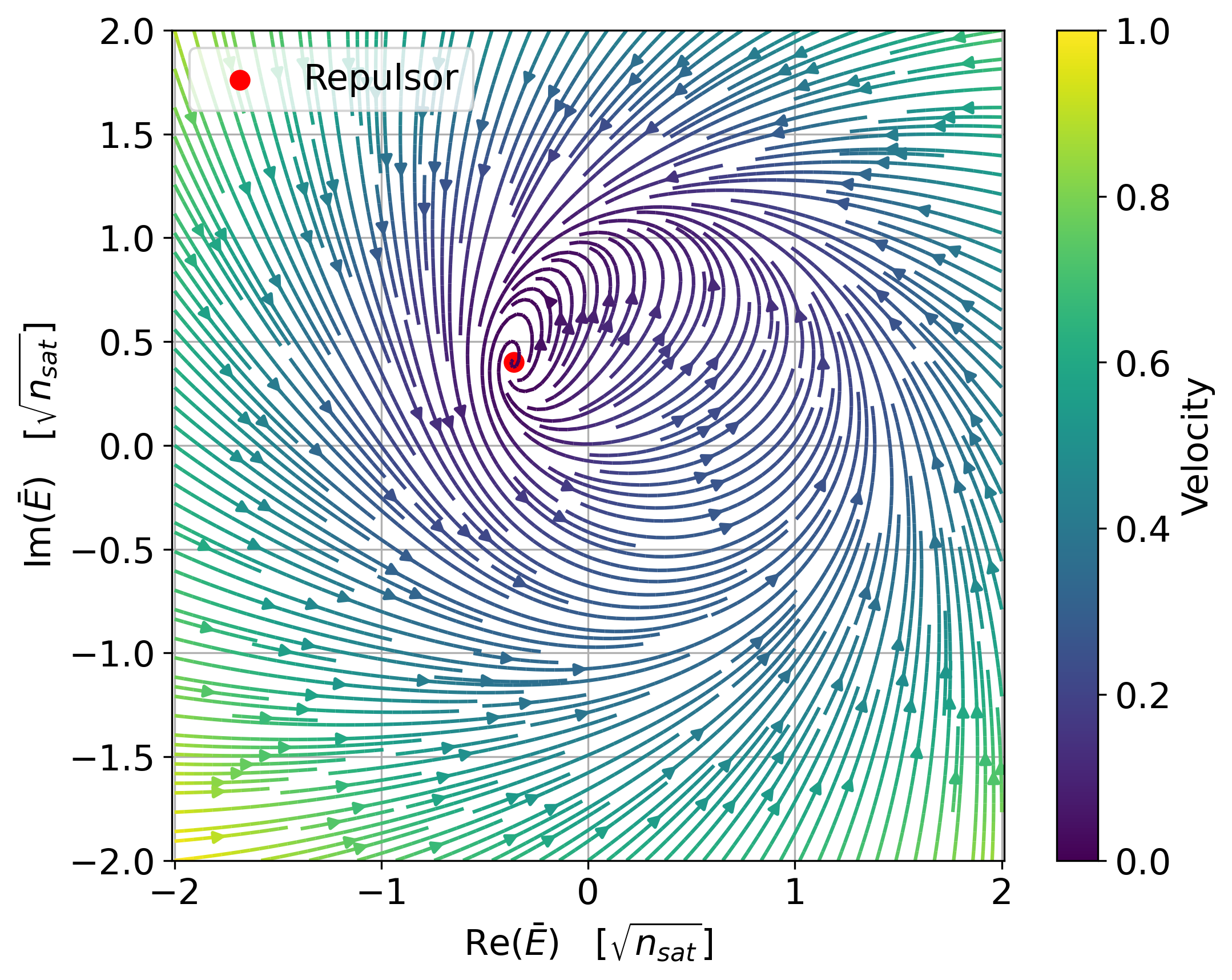}}
     \caption{\rev{Examples of vector flows of  $\bar{E}$ in the complex plane for $\tilde{P}=2$, strong detuning ($\tilde{\Delta}=0.3$) and no interactions ($\tilde{g}=0$), with decreasing values of $\tilde{E}_{coh}=[0.25,0.22]$ corresponding to gray dashed lines on the green curve in the left panel of Fig.\ref{graph: qualitative graphs cases 1 and 2}.  The panels show the formation of a limit-cycle from a single-solution region as $\tilde{E}_{coh}$ is decreased.} }
     \label{fig: limit-cycle formation at high Delta}
\end{figure*}

For relatively small $\Delta$ (red curves), the behavior is similar to the one of the $\Delta=0$ case. At low $P/\gamma$, there is a single stationary state solution with the cavity field intensity $I_{ss}$ monotonically growing with $I\rev{_{coh}}$. For large $P/\gamma$, multiple solutions are present, but only the uppermost one is dynamically stable \rev{with respect to spatially-uniform perturbations}.
As a main difference from the $\Delta=0$ case, the lower bound of the multi-solution region is no longer at $I\rev{_{coh}}=0$: this means that the high-$I_{ss}$ solution only exits above a threshold value of $I\rev{_{coh}}$. Below this value, the incident field is too weak to efficiently lock the cavity field, \rev{leading to a limit-cycle behavior of the steady state}.\\
\indent For large $\Delta$ (green curves) and for large $P/\gamma$ there is a single stationary state with the cavity field intensity $I_{ss}$ monotonically growing with $I\rev{_{coh}}$. However, its dynamical stability \rev{with respect to spatially-uniform perturbations} is guaranteed only for large enough values of $I_{ss}$, signaling again that the incident field can lock the cavity field only at sufficiently large amplitudes. Once again, below this value the stationary state is replaced by a limit cycle.\\
\indent These considerations on the stability of these solutions are illustrated in the flow diagrams shown in Fig.\ref{fig: limit-cycle formation}, which summarize the temporal dynamics of spatially uniform solutions $\bar{E}(\textbf{r},t)=\bar{E}(t)$ in the complex plane.
For relatively small $\Delta$, the two lower $I_{ss}$ solutions (red and yellow dots) in the multi-solution region are unstable and only the upper one (blue) is stable as visible in the upper-right and bottom-left panels. Then, for larger $E\rev{_{coh}}$ the two lower solutions merge and disappear (upper-left panel). For smaller $E\rev{_{coh}}$, instead, the middle and upper-$I_{ss}$ solutions merge leaving only the lower-$I_{ss}$ solution, which is however unstable: the system has no available stable stationary solution and the dynamics tends to a limit cycle  (bottom-right panel).\\
\indent Examples of the flow diagram in the large $\Delta$ case are shown in Fig.\ref{fig: limit-cycle formation at high Delta}: once again, the single stationary solution is stable at large $I\rev{_{coh}}$ (left) only, while at small $I\rev{_{coh}}$ (right) it turns unstable and is replaced by a limit cycle.

\begin{figure*}[htbp!]
\centering
\parbox[t]{0.49\textwidth}{%
    \includegraphics[width=0.85\linewidth]{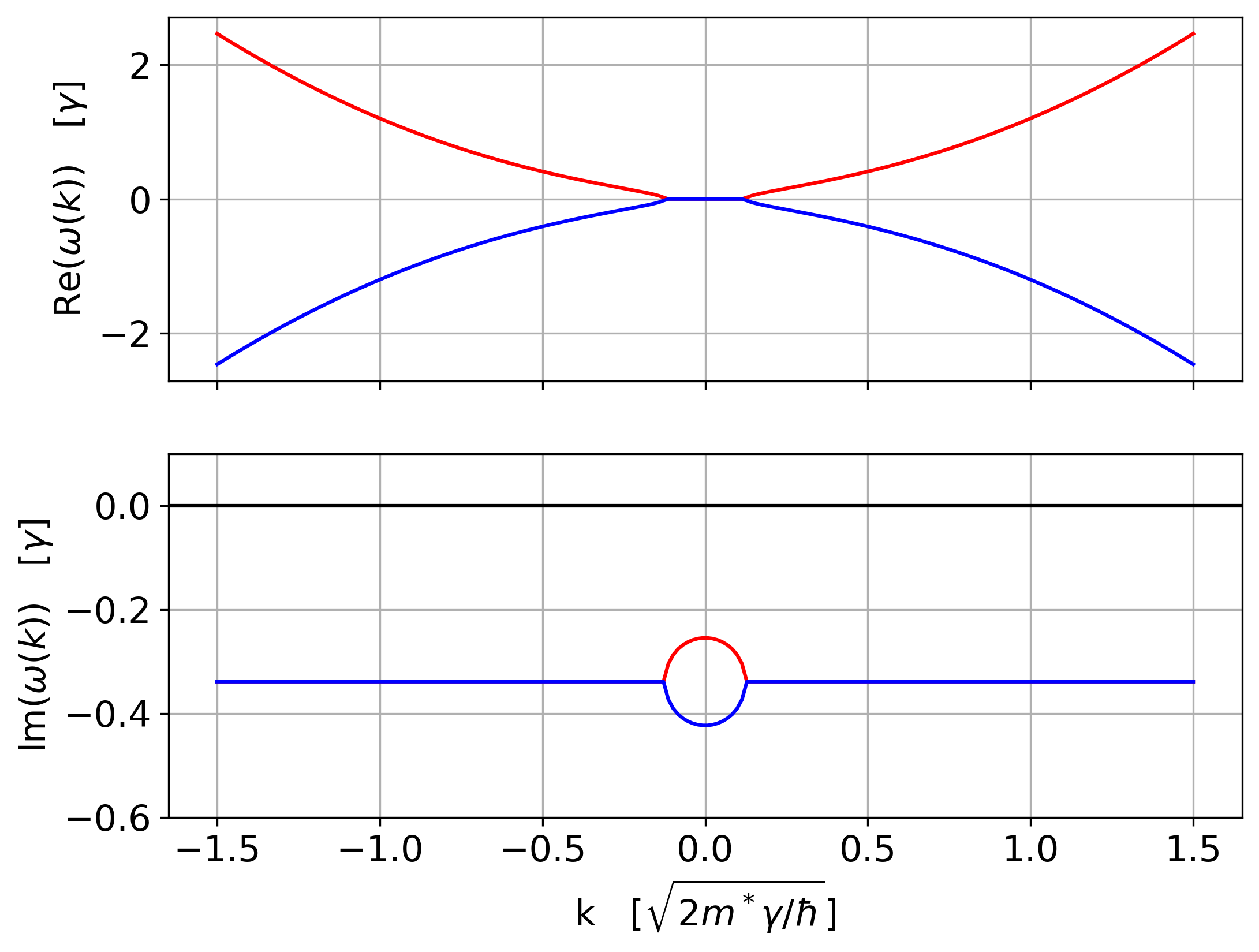} \\
    \includegraphics[width=0.85\linewidth]{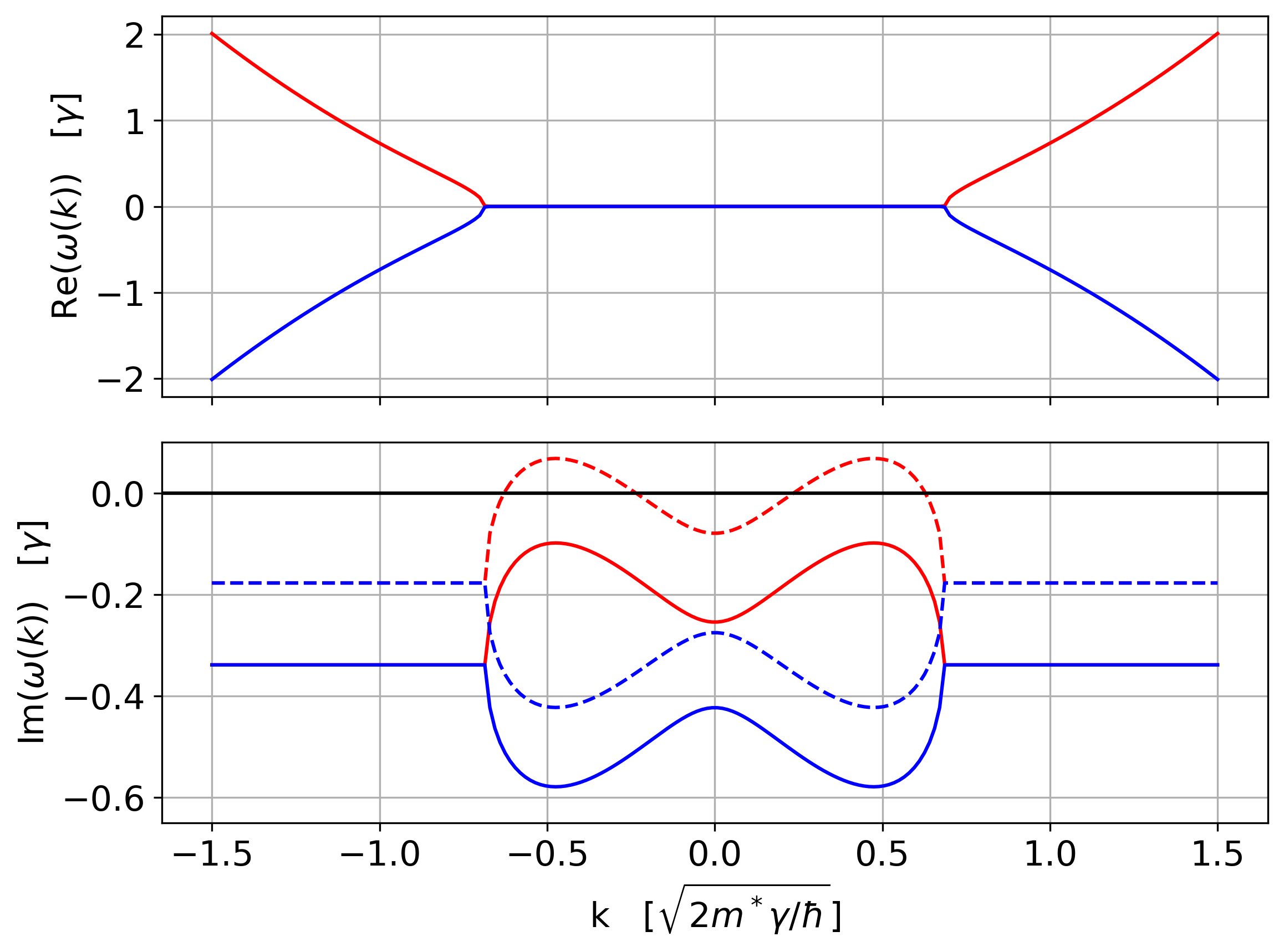}%
}       
\hfill
\parbox[t]{0.49\textwidth}{%
    \includegraphics[width=0.83\linewidth]{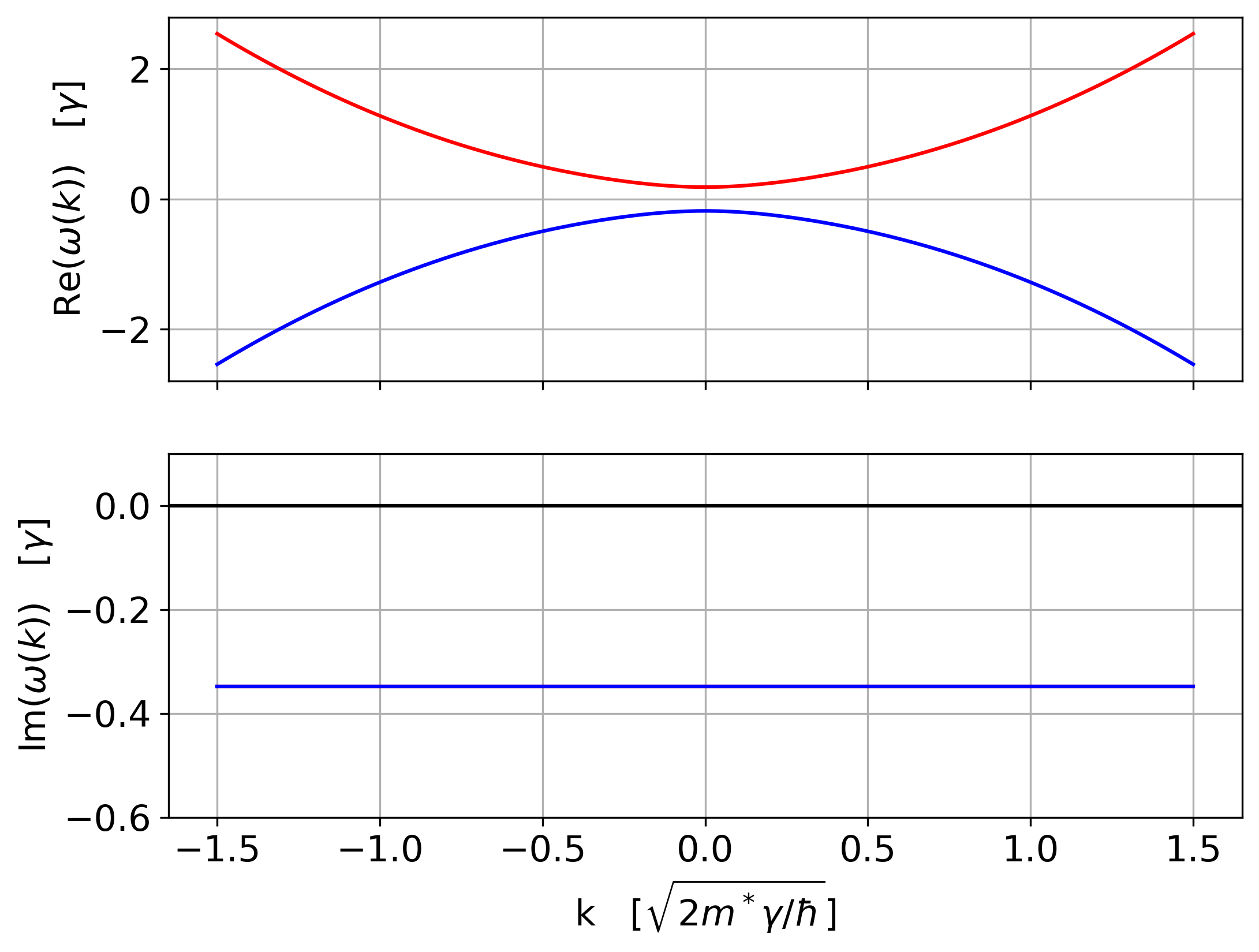} \\
    \includegraphics[width=0.83\linewidth]{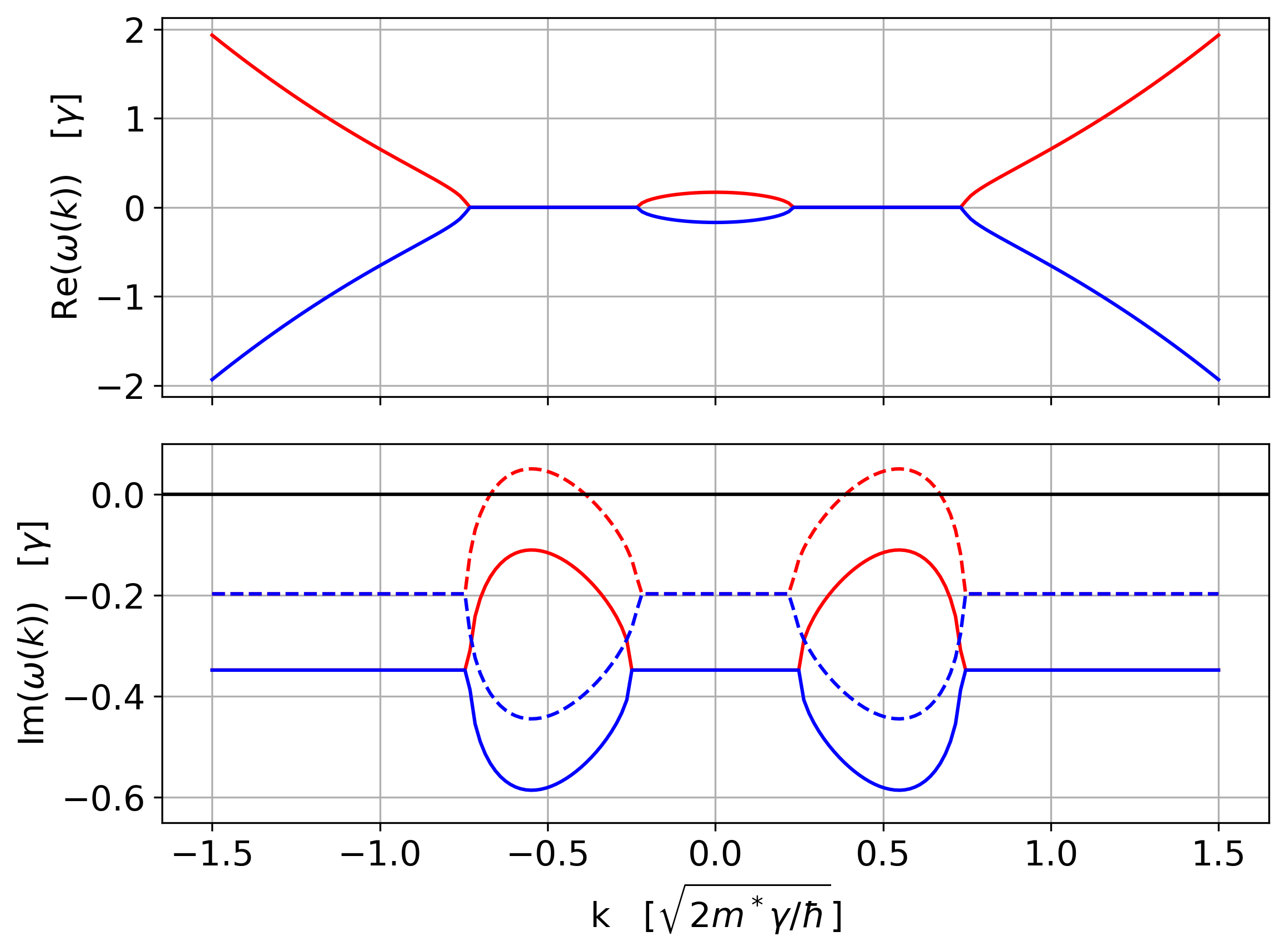}%
}
\caption{
\rev{Different types of Bogoliubov dispersion for the stable solution at $\tilde{P}=2$, fixed detuning $\tilde{\Delta}$ and no interactions ($\tilde{g}=0$). 
Left: absence of a real gap at $k = 0$, with the two examples marked by green points in Fig.~\ref{graph: phase diagram}: $\tilde{E}_{\mathrm{coh}}=0.3$, $\tilde{\Delta}=-0.225$ (top) and $\tilde{\Delta}=+0.225$ (bottom, solid lines). 
Right: emergence of a real gap at $k = 0$, with the two examples marked by blue points in Fig.\ref{graph: phase diagram}: $\tilde{E}_{\mathrm{coh}}=0.4$, $\tilde{\Delta}=-0.3$ (top) and $\tilde{\Delta}=+0.3$ (bottom, solid lines). 
Dashed curves in the imaginary part indicate the onset of instability at finite wavevectors $k \neq 0$, approaching the transition to a limit cycle; the corresponding parameters are marked by yellow points in Fig.\ref{graph: phase diagram}: left, $\tilde{E}_{\mathrm{coh}}=0.205$, $\tilde{\Delta}=0.225$; right, $\tilde{E}_{\mathrm{coh}}=0.275$, $\tilde{\Delta}=0.3$.}}

\label{fig:case2}
\end{figure*}

\begin{figure*}[htbp!]
    \centering
    \parbox[t]{0.49\textwidth}{%
    {\includegraphics[width= \linewidth]{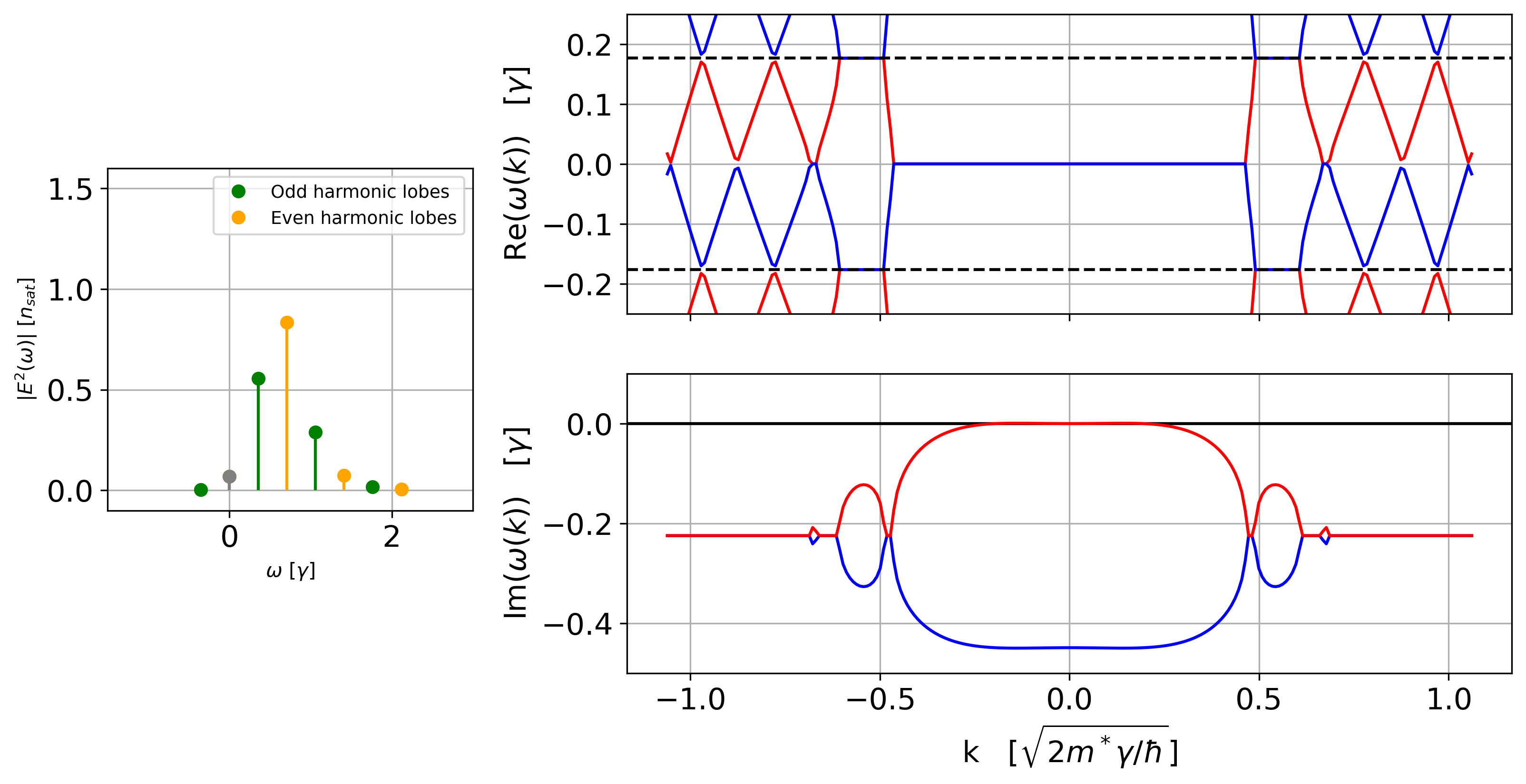}}
    }
    \hfill
    \parbox[t]{0.49\textwidth}{%
    \includegraphics[width= \linewidth]{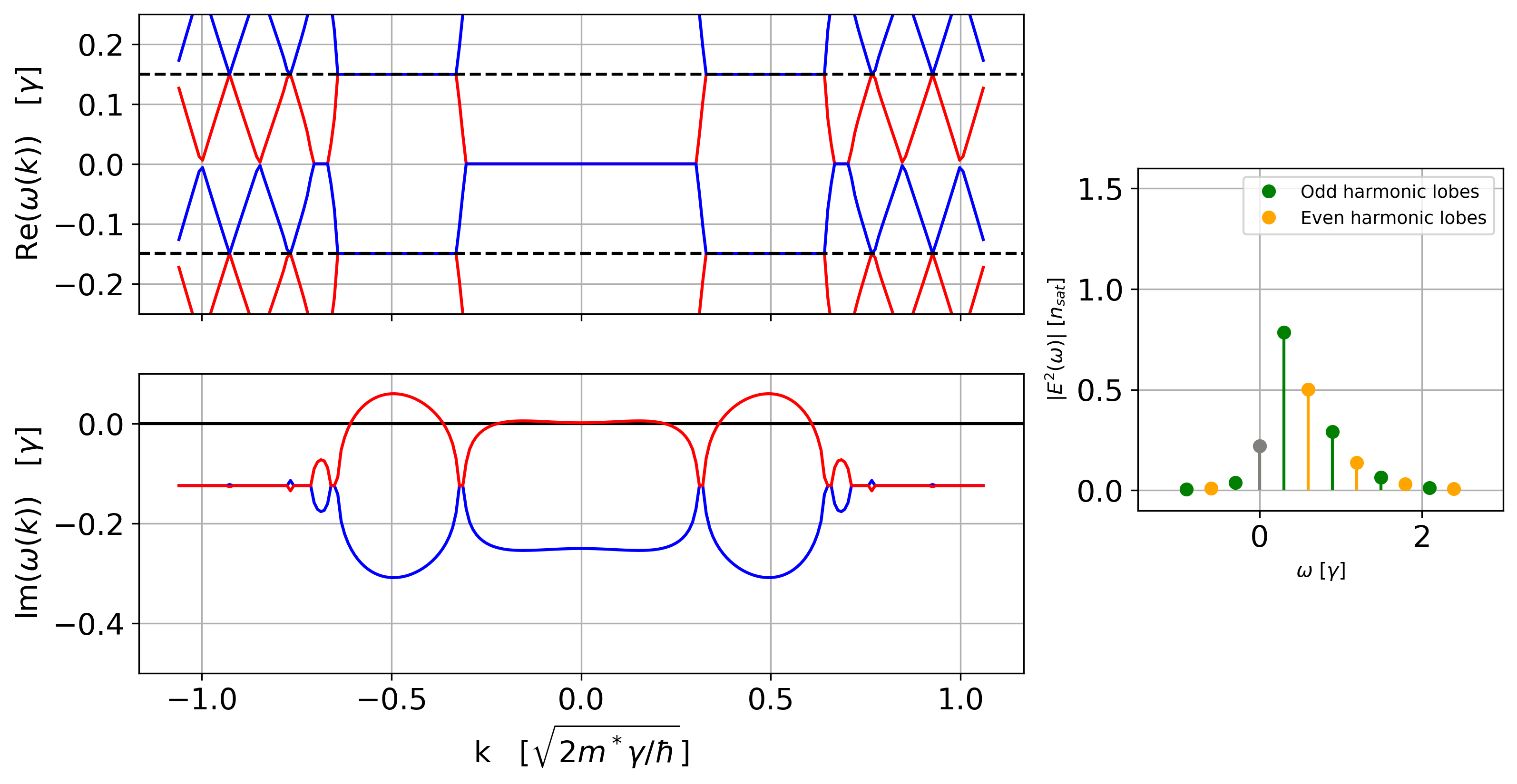}
    }
\caption{\rev{Examples of Floquet-Bogoliubov dispersion around a limit cycle at $\tilde{P}=2$, fixed detuning $\tilde{\Delta}$ and no interactions ($\tilde{g}=0$). A stable configuration (left) becomes unstable at finite $k$ (right) as the system approaches the transition to a stationary state. The corresponding parameters are marked by red points in Fig.\ref{graph: phase diagram}: $\tilde{\Delta}=0.375$, $\tilde{E}_{coh}=0.15$ (left) and $\tilde{E}_{coh}=0.25$ (right). The insets show the Fourier spectrum of $E_{ss}^{cyc}(t)^2$ along the limit cycle, whose period is such that $\tilde{\omega}_{ss}= 0.35 $ ($0.3$) in the left (right) panel. Alternating colors highlight the dominant harmonics: green marks components that couple to parametric resonances at the boundary of the Floquet Brillouin zone, while orange indicates components that couple to resonances at the center of the Floquet Brillouin zone.}}
    \label{Floquet general Einc}
\end{figure*}
\subsubsection{Bogoliubov \rev{dispersion of the collective excitations} around a stationary state}
\label{sec:Bogo_stationary}

These rich features reflect into different forms of the Bogoliubov dispersion of the collective excitations in the different cases. In this Section we will focus on collective excitations around stationary solutions, while in the next Section we will consider collective excitations around a limit cycle.

The Bogoliubov dispersion around a stationary state at relatively small $\Delta$ is shown in the left panels of Fig.\ref{fig:case2}. Similarly to the $\Delta=0$ case of Fig.\ref{graph: flow line case 1}, the gap at $k=0$ is a purely imaginary one and acquires a real part at large $k$.
As a key novelty, the $k^2$ kinetic energy term in \eqref{eq: dispersion relations} breaks the symmetry of the $\pm \Delta$ solutions, giving different shapes of the Bogoliubov dispersions at intermediate $k$ values in the two cases. The negative $\Delta<0$ case shown in the upper-left panel remains qualitatively similar to the $\Delta=0$ case throughout all $k$ values, the only difference being a shrinking of the central plateau. 
In the positive $\Delta>0$ case shown in the bottom-left panel, instead, the detuning $\Delta$ can be compensated by the kinetic energy term giving local maxima of the imaginary part around the points $k\simeq\sqrt{2m^*\Delta/\hbar} \neq 0$. 

It is interesting to note that the maximum of the imaginary part can cross to positive values at finite-$k$.
Analytical considerations reported in App.\ref{app:supersolid} show that this finite-$k$ instability occurs for the resonant modes around $k\simeq\sqrt{2m^*\Delta/\hbar}$  \rev{when}
\begin{equation}
    \frac{P}{1+|E_{ss}|^2/n\rev{_{sat}}}>\gamma\,,
\end{equation}
that is when gain saturation by the stationary field is not sufficient to suppress effective gain on the other modes. We highlight that this can only occur if $P>\gamma$ and that this threshold between finite-$k$ unstable and stable stationary states is the same of the lasing threshold in the same system in the absence of the coherent incident field $E\rev{_{coh}}$. Plugging in this formula the relation between $E_{ss}$ and the drive parameters, one obtains the condition
\begin{equation}
    \Delta>\frac{E\rev{_{coh}}}{\sqrt{n\rev{_{sat}}\left(\frac{P}{\gamma}-1\right)}}
\end{equation}
for the finite-$k$ instability (note that we have assumed from the beginning that $E\rev{_{coh}}$ is real and positive).
Explicit calculations show that, for decreasing $E\rev{_{coh}}$ at a given $\Delta$, the finite-$k$ instability appears before the onset of the limit-cycle instability of the uniform field. The dashed line in the bottom-left panel shows an example of Bogoliubov dispersion for a finite-$k$-unstable case: as expected, the finite-$k$ instability appears before the limit-cycle one at $k=0$.

In the presence of the finite-$k$ instability, the system does not admit any spatially uniform stable solution and tends to develop a spatial modulation along the $xy$ plane. The study of spatially inhomogeneous steady-state solutions was excluded from our treatment from the outset but we can conjecture~\cite{columbo2021unifying} that for suitable parameters the system might tend to a temporally stationary yet spatially periodic solution that spontaneously breaks the translational invariance. On the other hand, as the condensate phase is locked to the coherent drive, the $U(1)$ phase symmetry is explicitly broken. A complete investigation of this physics will be the subject of future work.

The case of a larger $\Delta$ is shown in the right panels of Fig.\ref{fig:case2}. As a main feature, the gap at $k=0$ may also contain a non-vanishing real part. This is due to the $\Delta^2$ contribution under the square root in \eqref{eq: dispersion relations} and can be physically understood as the system tending to oscillate at the natural cavity frequency $\omega_0$ rather than at the one imposed by the external drive $\omega\rev{_{coh}}$. Quite interestingly, it is indeed this frequency that is selected for the self-oscillation when the amplitude of the external drive is reduced and the system enters a limit-cycle behavior. 

Also in this relatively large-$\Delta$ case, the behavior at large $k$ is the same for $\pm \Delta$, but marked differences appear again for intermediate values of $k$. For negative $\Delta$, the real part is positive and grows smoothly and the imaginary part remains flat and featureless (upper-right panel). For positive $\Delta$, lobes appear in the imaginary part for growing $k$ with local maxima around the wavevectors $k\simeq\sqrt{2m^*\Delta/\hbar} \neq 0$ where the kinetic energy exactly compensates the detuning (bottom-right panel). Correspondingly to these lobes, the real part displays flat regions at $0$.
For suitable parameters, a positive value of the maximum imaginary part signals the onset of a modulational instability: an example of such finite-$k$ unstable Bogoliubov dispersion is shown as a dashed line in the bottom-right panel.

\subsubsection{Floquet-Bogoliubov \rev{dispersion of the collective excitations} around a limit cycle}
\label{sec:BogoFloquet2}

As discussed in Sec.\ref{sec:BogoFloquet}, the collective excitations around a limit cycle can be stroboscopically studied by monitoring the field at discrete times separated by the limit cycle period $T$. This requires taking the logarithm \eqref{eq:log} of the eigenvalues of the linearized propagator $U(T)$ for small perturbations around the limit cycle. 
Examples of the resulting dispersion curves are plotted in Fig.\ref{Floquet general Einc}. 

For a relatively weak coherent drive $E\rev{_{coh}}$ (left panel), the imaginary part is always negative. This illustrates the dynamical stability of the limit cycle orbit, that plays the role of an attractor. 
Furthermore, we observe that no gap is present for $k=0$ and the Bogoliubov dispersion displays for small $k$ the typical diffusive behavior of the Goldstone mode of driven-dissipative systems, with a flat and vanishing real part and a quadratic growth of the imaginary part towards negative values. \rev{Via a generalized version of the Goldstone theorem, this behavior is a direct consequence of the spontaneous selection of the starting position along the limit-cycle for the dynamics in each realization of the system, with different realizations generally corresponding to different, uncorrelated positions. As the field remains spatially uniform while evolving periodically along this limit cycle, this symmetry breaking can equivalently be interpreted as the usual spontaneous breaking of the 
$U(1)$ phase symmetry upon condensation.} In agreement with this picture, the Floquet-Bogoliubov dispersion continuously connects with the standard diffusive Goldstone mode found in the $E\rev{_{coh}}=0$ case~\cite{Wouters:PRL2007,claude2025observation}.\\
\indent As a consequence 
of the temporal periodicity of the limit cycle solution $E_{ss}^{cyc}(t)$, the Bogoliubov bands show a folding along the $\omega$-axis with periodicity $2\pi/T$ according to the Floquet-Brillouin zone picture. Around the crossing points between bands, this leads to the appearance of additional $k$-space regions where the Bogoliubov bands stick, giving a flat real part and lobes in the imaginary one.
\rev{Physically, these lobes originate from parametric instabilities induced by the periodic modulation during the limit-cycle dynamics. More precisely, the perturbation dynamics is driven by the periodic quantities $|E_{ss}^{cyc}(t)|^2$ and $E_{ss}^{cyc}(t)^2$, which appear in the diagonal and off-diagonal entries, respectively, of the time-dependent Bogoliubov matrix $M_{\mathbf{k}}(t)$ in (\ref{eq:matrixM}). As the off-diagonal entries are primarily responsible for coupling the Bogoliubov modes, it is mainly 
$E_{ss}^{cyc}(t)^2$ that act as the effective parametric drive responsible for the instability.}\\
\indent \rev{Typically, the most efficient coupling channel in parametric instabilities occurs via the well-known 2:1 resonance condition \cite{arnold1989mathematical}, where the frequency of the mode is half the frequency 
of the parametric modulation. This dominant process largely determines the growth of the lobes and, except for the central lobe at $k = 0$, 
allows a direct mapping of the newly formed lobes to the corresponding Fourier harmonics of $E_{ss}^{cyc}(t)^2$, with lobes emerging where the bands touch at the Floquet Brillouin zone boundaries (odd harmonics coupling) or at the center (even harmonics coupling). As shown in the external panels of Fig.\ref{Floquet general Einc}, the amplitude of higher Fourier harmonics quickly drops, which strongly suppresses both the height and the width in $k$ of the higher-$k$ lobes. An additional factor contributing to the reduction of the width in $k$ is the increasing slope of the kinetic energy at large $k$, which shifts the mode frequency rapidly away from that required for the fundamental parametric resonance.}\\
\indent \rev{For suitable parameters, e.g. a stronger coherent field}, the maximum of these lobes can cross beyond zero, so the limit cycle displays dynamical instabilities at finite $k$  \rev{(right panel of Fig.\ref{Floquet general Einc})}. This leads to a spatial modulation of the condensate and for suitable parameters might result in a spatially periodic steady-state. Differently from the case discussed in Sec.\ref{sec:Bogo_stationary} where only the translational symmetry was broken and the condensate phase remains locked to the coherent drive, two symmetries are here spontaneously broken: the translational symmetry as in a crystalline solid and the $U(1)$ phase symmetry of the condensate. In analogy to recent developments in ultracold atomic gases~\cite{recati2023supersolidity} and exciton-polariton fluids~\cite{Nigro:PRL2025,trypogeorgos2025emerging}, this novel state might then be considered as another candidate for a \textit{supersolid} state of light.
\rev{Whether this supersolid-like state corresponds to a traveling wave or exhibits features of a standing wave remains an open question. A definitive answer would require to go beyond the linearized dynamics and solve the full GPE \eqref{rotating frame field equation}, 
so to capture the complete spatio-temporal dynamics. This lies beyond the scope of the present work and will be addressed in future studies.}

\begin{figure}[h!]
{\includegraphics[width=\linewidth]{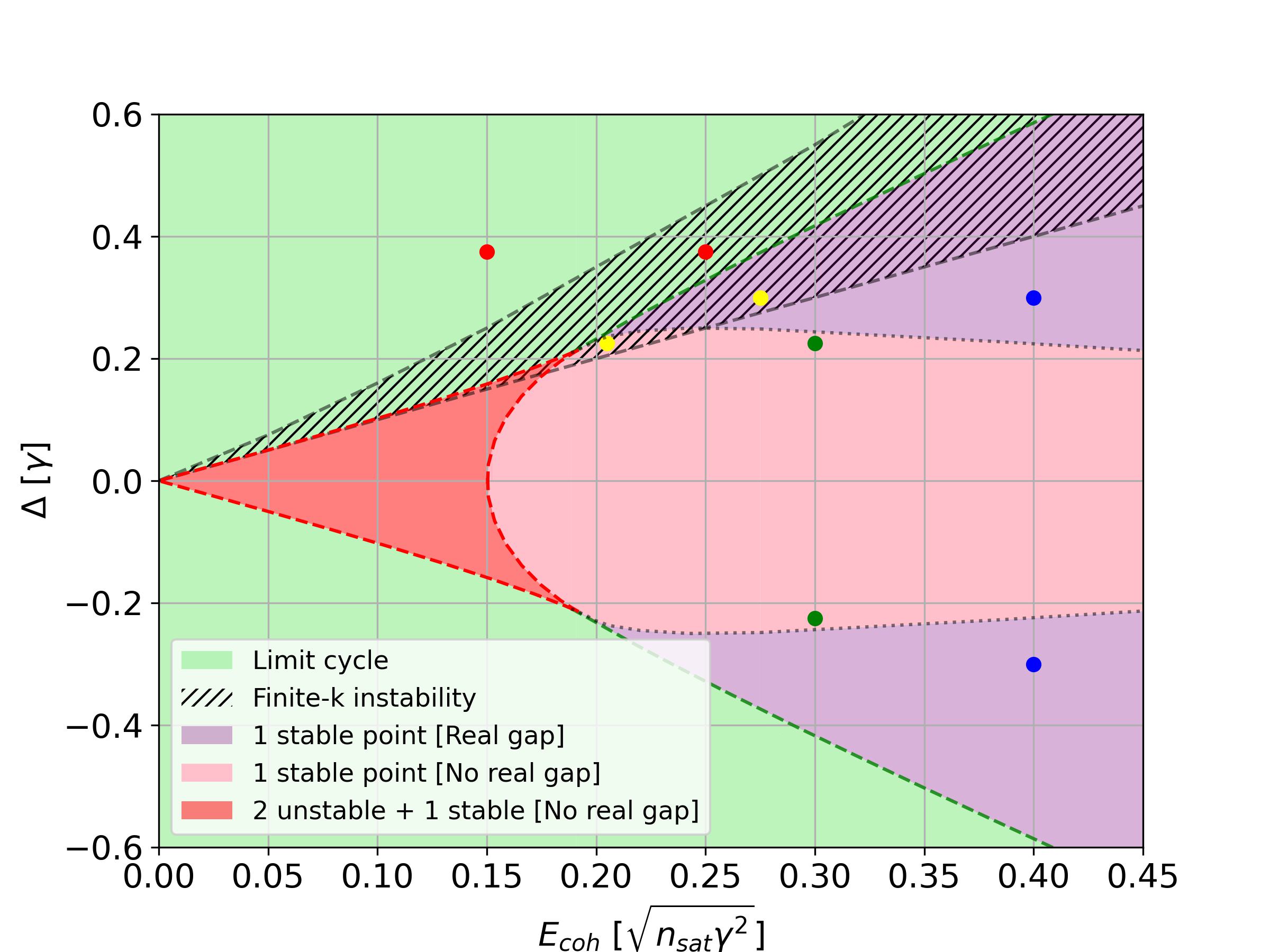}}
\caption{\rev{Phase diagram of the \rev{steady-state solutions} in the parameter space $(E\rev{_{coh}},\Delta)$ for $\tilde{P}=2$ and in absence of interactions ($\tilde{g}=0$). Colored points represent the parameter choices at which the Bogoliubov spectra in Fig.\ref{fig:case2}-\ref{Floquet general Einc} were calculated.}}
\label{graph: phase diagram}
\end{figure}

\subsection{Phase diagram: steady states vs limit cycles}
\label{sec: phase diagram}

We conclude the Section by summarizing the different regimes for a given value of the incoherent pump strength above threshold $P/\gamma>1$ into a single phase diagram \rev{\footnote{Numerical continuation of the bifurcation curves was performed using the MATLAB software package MatCont, see \rev{\cite{dhooge2008new}} }} as a function of the detuning $\Delta$ and the amplitude $E\rev{_{coh}}$ of the external coherent drive. An example of such phase diagram is shown in Fig.\ref{graph: phase diagram}.

The different colors correspond to different behaviors, namely stationary state vs. limit cycle and no real gap vs. real gap at $k=0$. The hatching indicates the regions where some finite-$k$ modes are unstable: here the system does not admit any spatially uniform stable solution and will develop a spatial modulation in the $xy$ plane, with the possibility of eventually reaching a spatially modulated steady state. 

As a general remark, we note that the coloring is symmetric under a change of the sign of $\Delta$: as mentioned above, the equation of motion for the uniform field component at $k=0$ are in fact complex conjugate for $\pm \Delta$. On the other hand, the hatching is non symmetric, reflecting the fact that the Bogoliubov dispersion  is strongly affected by a change in the sign of $\Delta$ and, as mentioned above, the finite-$k$ instability is only present on the $\Delta>0$ side. In specific, we can identify the following regions:

\begin{enumerate}[label=(\Alph*)]

\item In the red and pink areas, the system tends to a unique stable stationary state where the condensate phase is locked to the coherent drive one. The Bogoliubov dispersion of collective excitations features a purely imaginary gap at $k=0$. The red region indicates the multi-solution region with two unstable stationary states and one stable. In the pink region only one stable stationary solution exists. In agreement with Fig.\ref{graph: qualitative graphs cases 1 and 2} the boundary of the red region reaches for $\Delta=0$ the $E\rev{_{coh}}=0$ point. 
    
\item In the violet area, a unique stable stationary state exists but the gap in the Bogoliubov dispersion also displays a finite real part at $k=0$.  Interestingly, one can see from \eqref{eq: dispersion relations} that for $E\rev{_{coh}} \to \infty$ the boundary of this real-gap region asymptotically tends to $\Delta \to 0$: any small detuning is able to open a real gap if a strong enough coherent field is present. 

\item In the green area, the system tends to a limit cycle and the Floquet-Bogoliubov dispersion shows a gapless Goldstone branch as a consequence of the spontaneously broken $U(1)$ symmetry associated to the condensate phase. \\ The finite-$k$ instability in the hatched green region is a candidate for realizing an optical analog of a supersolid state~\cite{recati2023supersolidity} where both the spatial translation and the condensate phase symmetries are simultaneously broken. It is interesting to note that, even though the coherent drive is unable to effectively lock the condensate phase, its very presence favors~\cite{columbo2021unifying} the onset of the finite-$k$ instability. As compared to recent observations of polariton supersolidity~\cite{Nigro:PRL2025,trypogeorgos2025emerging}, the configuration considered in this work does not require multiple photonic branches.

\end{enumerate}

\rev{In the regime of a weak coherent drive $E_{coh}$, the boundary separating the phase-locked stationary state from the limit-cycle region displays a shape that qualitatively resembles the edge of a \textit{1:1 Arnold tongue} (the only locking condition present in our model involving a rotating-wave approximation) of a weakly forced nonlinear oscillator \cite{kuznetsov1998elements,golden2021arnold}. 
In the standard synchronization scenario, this locking region opens around zero detuning with a width that grows linearly with the drive amplitude. Similarly, in our phase diagram, for weak $E_{coh}$ the stationary phase-locked solution exists only within a finite interval of detuning that widens as $E_{coh}$ increases. This analogy provides useful intuition: the coherent field acts as an external oscillator that can synchronize the intrinsic condensate dynamics within a detuning window controlled by its amplitude.}

 Further light on the physics is obtained by specifically looking at the nature of the \rev{bifurcations~\cite{kuznetsov1998elements}
 } occurring at the transitions separating the different behaviors.

\begin{itemize}

\item The transition from one stable stationary solution in the purple region to a limit cycle in the green region is an example of \rev{\textit{Hopf bifurcation} 
\cite{kuznetsov1998elements}}. Approaching the boundary of the stable region, the (negative) imaginary part of the $k=0$ Bogoliubov mode around the stable stationary state grows towards zero until the stable stationary state transforms into an unstable point surrounded by a stable limit cycle as illustrated in the right panels of Fig.\ref{fig: limit-cycle formation}. As the radius of the limit cycle grows from zero starting from the Hopf bifurcation point, we can classify this phenomenon as a kind of second-order phase transition, associated to the spontaneous breaking of a $U(1)$-like symmetry related to the condensate phase as discussed in Sec.\ref{sec:BogoFloquet2}. Note how in the $\Delta>0$ region, this transition is preceded on both its sides by the finite-$k$ instability towards a spatially modulated state.
    
\item In the transition from the multi-solution red region to the green limit-cycle region illustrated in the left-bottom and right-bottom panels of Fig.~\ref{fig: limit-cycle formation}, the unstable low-$I_{ss}$ intensity solution remains unaffected, while the stable high-$I_{ss}$ and unstable intermediate-$I_{ss}$ solutions collide and disappear. They are replaced by a limit cycle with non-zero radius, \rev{in a process known in the literature as a \textit{saddle-node homoclinic bifurcation}~\cite{kuznetsov1998elements}.} 
As the transition to the limit cycle is approached, the (negative) imaginary gap of the Bogoliubov dispersion at $k=0$ closes without any real part. In the $\Delta>0$ region, this transition is preceded by a finite-$k$ instability towards a spatially modulated state, signaled by a smooth change of the Bogoliubov dispersion in which the maximum of the imaginary part crosses above $0$.

\item In the transition from the \rev{multi-solution} red region to \rev{one stable stationary solution with no real gap in} the pink region, the two unstable solutions at low and intermediate $I_{ss}$ undergo a \rev{\textit{saddle--node bifurcation} \cite{kuznetsov1998elements}}, where they collide and annihilate each other, as illustrated in Fig.~\ref{fig: limit-cycle formation}. The locus of these saddle--node points defines the curved lateral boundary of the red region. This transition has no impact on the Bogoliubov dispersion relations of the remaining stationary state, whose imaginary part remains negative for all $k$.

\begin{figure*}[htbp!]
    \centering
    \parbox[t]{0.485\textwidth}{%
    {\includegraphics[width=\linewidth]{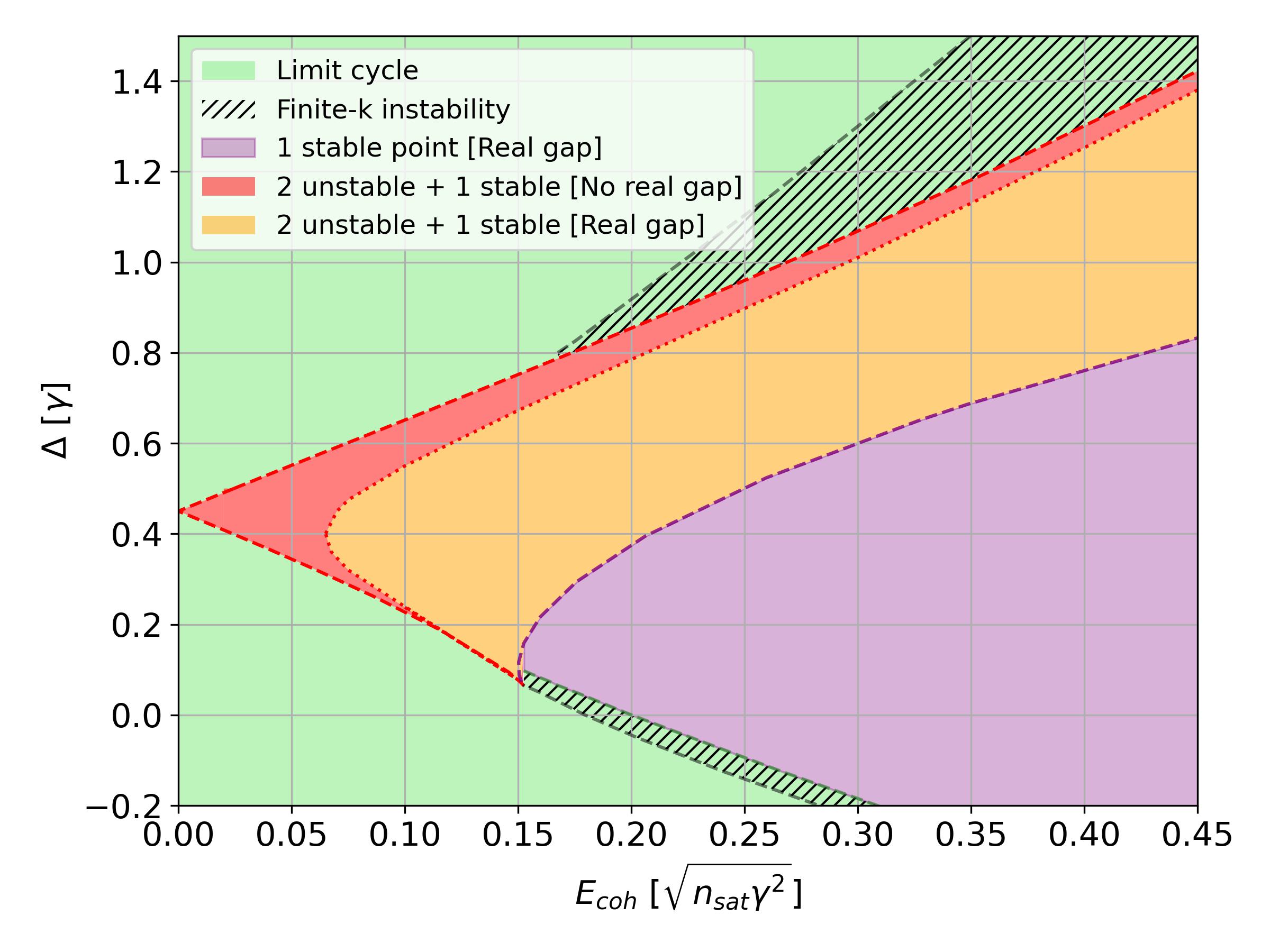}}
    }
    \parbox[t]{0.49\textwidth}{%
    \includegraphics[width=\linewidth]{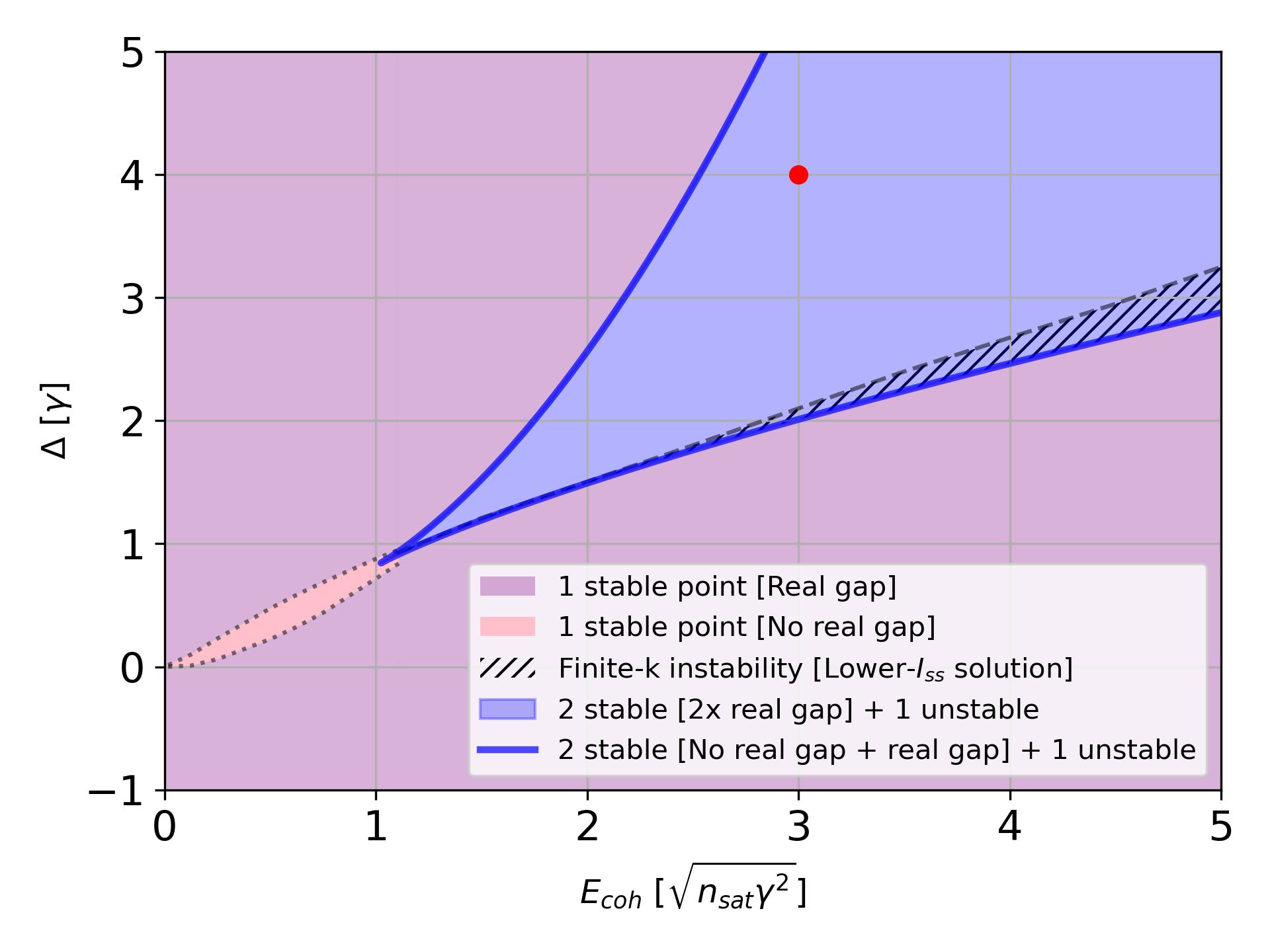}
    }
\caption{\rev{Left: Phase diagram of the steady-state solutions in the parameter space $(E\rev{_{coh}}, \Delta$) for $\tilde{P}=2$ and $\tilde{g} = 0.45$. Right: Phase diagram of the steady-state solutions in the parameter space $(E_{coh},\Delta)$, for $\tilde{P}=0.5$ and  $\tilde{g}=0.15$. 
The red point marks the parameter choice at which the vector flow shown in Fig.\ref{graph: multistable flow} was calculated.}}
\label{graph: phase diagram case 3 monostable}
\end{figure*}

\item \rev{The point where the two saddle-node bifurcation curves -- which by themselves would form a \textit{cusp point} \cite{kuznetsov1998elements} -- meet the Hopf bifurcation line resembles the features of what is called in the literature a \textit{Bogdanov--Takens (BT) bifurcation}~ \cite{kuznetsov1998elements}. In particular, approaching this point from the red region, the three steady-state solutions are expected to merge. This follows from the fact that along the red–green boundary the attractor and the saddle collide and disappear, while along the red–pink boundary the same occurs for the repulsor and the saddle. It is also worth noting that the limit cycle emerging at the red–green transition has a finite amplitude, whereas along the purple–green boundary it appears with vanishing amplitude. This suggests that, when following the phase boundaries toward the critical point, the apparent amplitude of the limit cycle changes abruptly, although a continuous connection exists when following the limit cycle along its continuation line at fixed period in parameter space.}

\end{itemize}

\section{Effect of interactions ($g \neq 0$)}

\label{sec:Case3}

In this most general case with a finite interaction constant $g$, the stationary state can be found by including the corresponding term in \eqref{steady state equation}. This leads to the relation
\begin{equation}
    I\rev{_{coh}}=I_{ss}\left((\Delta-gI_{ss})^2+\frac{1}{4}\left(\frac{P}{1+I_{ss}/n\rev{_{sat}}}-\gamma\right)^2\right)
    \label{eq: intensities case 3}
\end{equation}
between the incident $I\rev{_{coh}}$ and stationary $I_{ss}$ intensity, which can be straightforwardly reduced to a quintic polynomial equation. The phase difference between $E_{ss}$ and $E\rev{_{coh}}$ is now given by 

\begin{multline}
    \Delta\phi_{E_{ss},E\rev{_{coh}}}=\frac{\pi}{2}+ \\ + \arctan\left[\frac{1}{2(\Delta-g|E_{ss}|^2)}
     \left(\frac{P}{1+|E_{ss}|^2/n\rev{_{sat}}}-\gamma\right)\right]\,.
\end{multline}
Looking at \eqref{eq: intensities case 3}, one notices that a simultaneous reversal of the sign of both $\Delta$ and $g$ gives a complex-conjugate solution for $E_{ss}$ and leaves the intensity unchanged. As in the previous $g=0$ case, the Bogoliubov dispersion \eqref{eq: dispersion relations} is instead affected by this transformation. In what follows, we will focus for definiteness on the $g>0$ case. A numerical exploration of the stationary equation \eqref{eq: intensities case 3} for different choices of parameters suggests the following \rev{two} main regimes.\\
\indent The phase diagram plotted in the left panel of Fig.\ref{graph: phase diagram case 3 monostable} refers to the $P>\gamma$ regime with a relatively large value of the effective nonlinear parameter $g n\rev{_{sat}} \,(P-\gamma)/\gamma$.
For this choice of parameters, the multiple solutions of \eqref{eq: intensities case 3} involve a single stable high-$I_{ss}$ solution and two unstable ones at lower $I_{ss}$, very similarly to the $g=0$ case.
The stationary-state region is surrounded by a limit-cycle one. Interestingly, the tip of the stationary-state region is now located at a blue-shifted $\Delta=gn\rev{_{sat}} \,(P-\gamma)/\gamma$ as a consequence of the interaction term. Again, there are regions with a purely imaginary gap in the Bogoliubov dispersion  at $k=0$ and regions where this gap also displays a finite and positive real part.
Finally, we emphasize that also in this case finite-$k$ instability regions exist for the limit-cycle solutions, both in the vicinity of the transition to the multi-solution region and close to the Hopf-like transition toward the single-solution regime. However, for this chosen value of $g$, no region showing a stationary state with finite-$k$ instability was found. This is due to the presence of the large $2g|E_{ss}|^2$ term in \eqref{eq: dispersion relations} which shifts the dispersion countering the effect of a positive detuning and preventing the occurrence of finite-$k$ maxima with positive imaginary part; for smaller values of $g n\rev{_{sat}}\,(P-\gamma)/\gamma$, finite-$k$ instabilities become again possible also for the stationary states, as in the 
$g=0$ case.\\
\begin{figure}[ht]
{\includegraphics[width=0.55\textwidth]{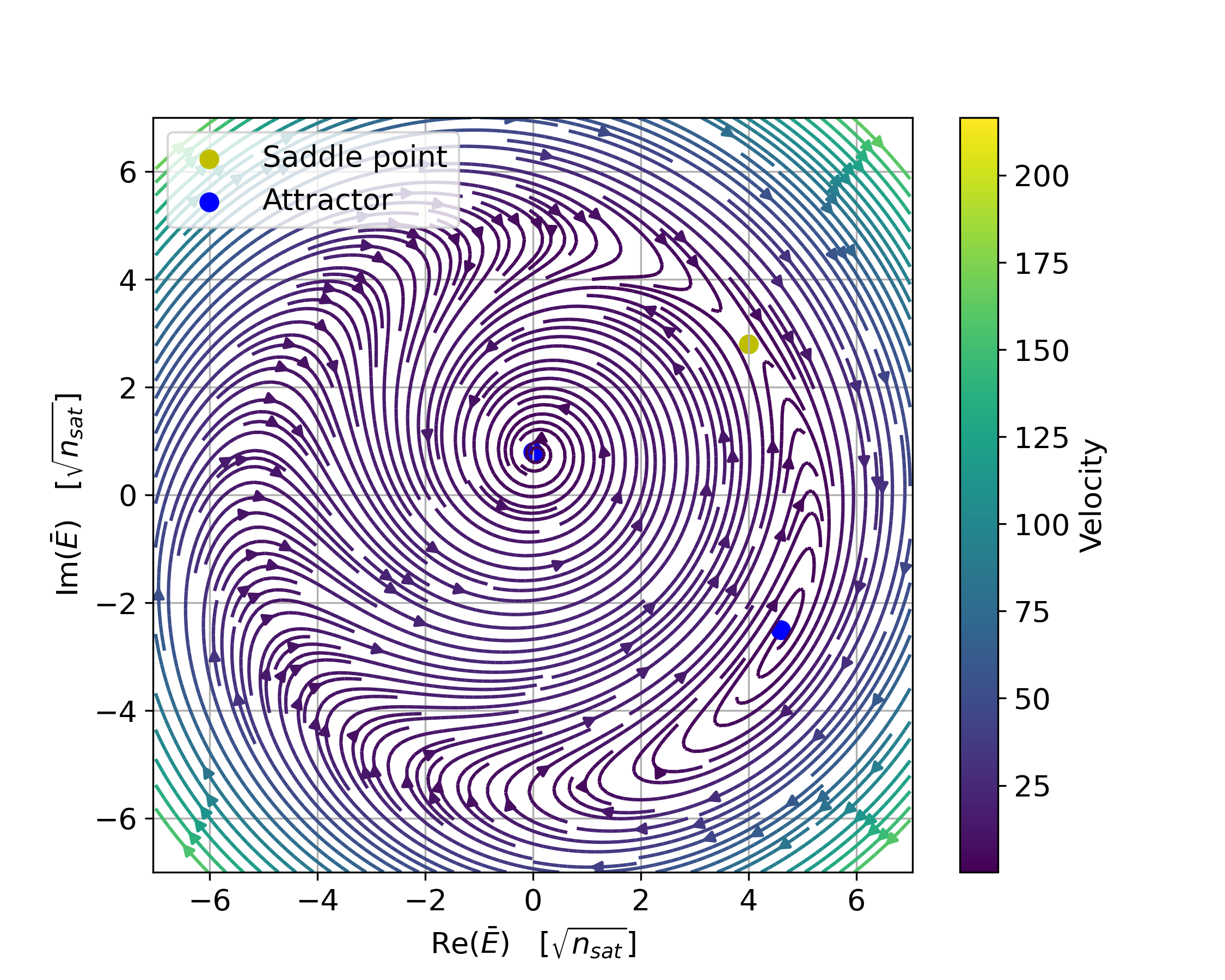}}
\caption{\rev{
Example of vector flow of $\bar{E}$ in the complex plane, corresponding to the red point shown in Fig.\ref{graph: phase diagram case 3 monostable}: $\tilde{P}=0.5$, strong detuning ($\tilde{\Delta}=4$) and $\tilde{g}=0.15$, at $\tilde{E}_{coh}=3$.}}
\label{graph: multistable flow}
\end{figure}
\indent For $P<\gamma$, the physics is reminiscent of the one of a coherently driven anharmonic oscillator, theoretically predicted in \cite{Carusotto:PRL2004} and experimentally observed in \cite{Claude:PRL2022}: the weak strength of the incoherent pump makes so that the cavity field is dominantly determined by the coherent drive to which it stays locked and the effect of the incoherent pump reduces to an effective reduction of the loss rate to $\gamma-P$.
Thanks to the intensity-dependent shift of the cavity frequency, the stationary state intensity \eqref{eq: intensities case 3} can display bistability effects for $\Delta>0$, with two stable solution at high- and low-$I_{ss}$ and a dynamically unstable intermediate-$I_{ss}$ one: an example of flow pattern for such a configuration is shown in Fig.\ref{graph: multistable flow}. \\
\indent The phase diagram is shown in the right panel of Fig.\ref{graph: phase diagram case 3 monostable}: \rev{since} the condensate phase is locked to the coherent drive, no limit-cycle region is present. 
Except for a thin region on the edges of the bistability region, the $k=0$ gap in the Bogoliubov dispersion has a finite real part for both stable solutions. The real gap vanishes only in the vicinity of the lower-$E\rev{_{coh}}$ boundary for the high-$I_{ss}$ solution, where this solution subsequently disappears; conversely at the higher-$E\rev{_{coh}}$ boundary the real gap closes for the low-$I_{ss}$ solution, just before it also ceases to exist. In analogy to the $P=0$ case~\cite{Carusotto:PRL2004}, the former boundary corresponds to the sonic behavior of the Bogoliubov dispersion. Finally, a region of finite $k$ instability is observed near the higher-$E\rev{_{coh}}$ boundary of the light blue region and only inside it: as in the $P=0$ case~\cite{Carusotto:PRL2004}, this occurs only to the Bogoliubov dispersion relations of the low-$I_{ss}$ solution.
\section{Conclusions}
\label{sec:Conclu}

\rev{In this work, we have developed a generic model of a driven-dissipative condensate under simultaneous incoherent pumping and coherent phase-locking drive to analyze the gap opening in the dispersion of its collective excitations (Sec.~\ref{sec:Model}). In the absence of inter-particle interactions, except for gain saturation, the condensate phase is efficiently locked when the coherent drive is sufficiently strong and near the natural cavity frequency: the dispersion of the excitation modes associated with condensation acquires an energy gap which, in addition to its imaginary part, can also exhibit a finite real part for increasing detunings (Sec.~\ref{sec:Case2} 1--2). For even larger detunings, the coherent field can no longer lock the condensate phase, which then evolves \rev{at a \textit{dynamically} chosen frequency} with a \textit{spontaneously} selected initial U(1) phase. \rev{This spontaneous symmetry breaking restores the gapless Goldstone mode, now displaying a novel Bogoliubov-Floquet dispersion with band foldings around the Floquet-Brillouin zone (Sec.~\ref{sec:Case2} 3).} \\
\indent Together, these behaviors define the phase diagram constructed in the plane of coherent drive amplitude and frequency (Sec.~\ref{sec: phase diagram}). Within this diagram, we further identify regimes featuring finite-wavevector dynamical instabilities that lead to spatial modulations of the condensate. The potential for stabilizing an optical analog of a supersolid state, which simultaneously exhibits phase coherence and a spatially modulated intensity profile, is a natural direction for future investigation.\\
\indent When inter-particle interactions are included, the dynamics become significantly more complex, with an interplay between condensation induced by the incoherent pump and bistability effects under the coherent drive. The phase diagram is illustrated in two relevant regimes corresponding to pumping strengths below and above the gain--loss threshold, revealing a range of behaviors reminiscent of the non-interacting case (Sec.~\ref{sec:Case3}).\\
\indent Although developed for a minimal condensation model, these qualitative conclusions are broadly applicable to generic condensates, optical parametric oscillators, and laser devices in spatially extended configurations, whether purely spatial~\cite{contractor2022scalable} or spatio-temporal~\cite{Lugiato:Varenna}. In particular, our model provides a clear theoretical and physical framework for interpreting recent experimental observations of the collective excitations of parametrically pumped exciton-polariton condensates in semiconductor microcavities~\cite{claude2025observation}.}

\begin{acknowledgments}

The research reported in the work was carried out by E.S. and G.A.P.S. as a part of their Quantum Optics exam at the Master in Physics of Trento University. 
I.C. acknowledges continuous exchanges with Alberto Bramati and Michiel Wouters, as well as financial support from: Provincia Autonoma di Trento (PAT); the Q@TN Initiative; the National Quantum Science and Technology Institute through  the PNRR MUR project under Grant PE0000023-NQSTI, co-funded by the European Union -- NextGeneration EU.

\end{acknowledgments}

\newpage
\nocite{*}
\bibliography{references}

\newpage
\appendix

\section{Analytical considerations on multiple solutions in the non-interacting case ($g = 0$)}
\label{app:ss_g=0}
Analytical conditions for the existence of multiple solutions at a given $I\rev{_{coh}}$ can be obtained studying the sign of the derivative
\begin{align}
    \frac{dI\rev{_{coh}}}{dI_{ss}}&=\left(\Delta^2+\frac{1}{4}\left(\frac{P}{1+I_{ss}/n\rev{_{sat}}}-\gamma\right)^2\right) + \notag \\ &+\frac{I_{ss}P}{2n\rev{_{sat}}(1+I_{ss}/n\rev{_{sat}})^2}\left(\gamma-\frac{P}{1+I_{ss}/n\rev{_{sat}}}\right)
    \label{eq: stability condition case 2}
\end{align}
For $\gamma>P$, the derivative is always positive and a single solution is present. However, in full generality, the condition $dI\rev{_{coh}}/dI_{ss}>0$ can be recast in polynomial form. Defining $x=1+I_{ss}/n\rev{_{sat}}>1$, the condition becomes:
\begin{equation}
    f(x)=x^3-\overbrace{\frac{P(P+2\gamma)}{\gamma^2+4\Delta^2}}^{C}x+\overbrace{\frac{2P^2}{\gamma^2+4\Delta^2}}^D>0\,.
    \label{eq: stability condition case 2 recast}
\end{equation}
As $df/dx = 3x^2 - C$, the function has only one minimum in $x_{min}=\sqrt{C/3}$. Since $f(1)>0$, the solution is unique if and only if:
\begin{equation}
    \{x_{min}<1\}\text{\quad OR \quad } \{ f(x_{min})>0\quad \text{AND} \quad x_{min}>1\}\,,
    \label{eq: stability condition case 2 final}
\end{equation}
otherwise the system displays multiple solutions in a region whose boundaries are found solving $f(x)=0$ for $x>1$ with the cubic formula.

\section{Analytical condition for finite-$k$ instability in the non-interacting case ($g = 0$)}
\label{app:supersolid}

Considering positive $\Delta>0$ detuning, the dispersion relations plotted in Fig.\ref{fig:case2} feature the possibility of a finite-$k$ instability of the stationary states and of the limit cycles. Here, we show that one of the curves delimiting such region is a straight line that can be analytically determined.

Considering (\ref{eq: steady state case 2}) and (\ref{eq: dispersion relations}) for $g=0$, we study for $\Delta>0$ the behavior of the maxima of the upper band in the imaginary part of the dispersion relations. From (\ref{eq: dispersion relations}), the maximum of the imaginary part is at $\hbar k^2/ (2m^{*})=\Delta$ and has
\begin{equation}
    \max[\textrm{Im}(\omega_{+})]=\frac{i}{2}\left(\frac{P}{1+|E_{ss}|^2/n\rev{_{sat}}}-\gamma\right)
\end{equation}
Now, the instability at finite $k$ arises when the term in parenthesis is positive, the threshold being at 
\begin{equation}
    |E_{ss}|^2 = n\rev{_{sat}}\left(\frac{P}{\gamma}-1\right)\,.
    \label{eq:threshold}
\end{equation}
Substituting this condition in (\ref{eq: steady state case 2}) we find that the threshold condition imposes that $\Delta\,E_{ss}=iE\rev{_{coh}}$. Since $\Delta>0$ and $E_{ss}$ is fixed by \eqref{eq:threshold}, the only possibility for this to occur is that $E_{ss}$ is rotated by 90\textdegree compared to $E\rev{_{coh}}$ in the complex plane (in this paper, we assumed $E\rev{_{coh}}$ real and positive and, thus, $E_{ss}$ is purely imaginary on the positive side of the imaginary axis) and that the parameters satisfy:
\begin{equation}
    \Delta=\frac{E\rev{_{coh}}}{\sqrt{n\rev{_{sat}}\left(\frac{P}{\gamma}-1\right)}}
\end{equation}

 We remark that this line sets the boundary only in stationary state regions and not in the limit-cycles one.

Referring to Fig.\ref{graph: phase diagram}, we also note that, starting from the stationary-state side, at fixed $E\rev{_{coh}}$ we obtain the region with finite-$k$ instability by increasing $\Delta$, while, at fixed $\Delta$, this instability is found by decreasing $E\rev{_{coh}}$.

\end{document}